**Life-Cycle Effects of Comprehensive Sex Education**

Volha Lazuka* and Annika Elwert**

**Abstract**

Using the introduction of comprehensive sex education in Sweden as a natural experiment, we explore how educational curricula can shape social norms and impact personal well-being. Inspired by liberal values, the curriculum taught more than just biology. It instilled lessons on abstinence, rational family planning, and the importance of taking social responsibility for personal decisions. We find that the reform successfully addressed its intended outcomes, reducing unwanted pregnancies, leading to fertility postponements and increasing female labor force participation. The findings suggest that social norms, internalized through school-based sex education, persistently affect people's outcomes in significant ways.

* vola@sam.sdu.dk, Department of Economics, University of Southern Denmark, Campusvej 55, M-5230 Odense; Department of Economics, Lund University, and IZA; +45-93 50 76 81.

** annika.elwert@soc.lu.se, Department of Sociology, Lund University.

## 1 Introduction

Institutions such as schools shape norms and identities, which have consequences for students' economic well-being and gender equality (Kranton 2016; Akerlof and Kranton 2010). In many countries, sex education has long been integrated into the compulsory school curriculum. The content and scope of sex education vary, ranging from abstinence-only approaches (which are becoming increasingly popular in many US states) to comprehensive programs that include ethical components addressing healthy relationships, life skills, and social norms (as seen in Sweden and other countries). Recent experimental studies have shown that schools have the ability to influence students' social and gender attitudes (Dhar, Jain, and Jayachandran 2022; Cappelen et al. 2019), particularly through comprehensive sexuality education (DeGue et al. 2021; Oringanje et al. 2016). However, the limited duration of these experiments prevents an assessment of the effects of social norms throughout the life cycle.

In this study, we investigate how the transmission of social norms—by means of a sex education reform in primary school—could affect the life choices and economic well-being of men and women. Sweden became a global forerunner in implementing comprehensive sex education in the 1940s–1950s. The implementation was a response to the population crisis of the 1930s and aimed at easing the burden of unwanted childbearing for poor families without lowering overall fertility. The reform promoted abstinence and rational choice in sexual and family matters but was not limited to it. It also served as a pioneering social intervention during the building of the social democratic welfare state, aiming to educate adolescents to become individually and collectively responsible citizens. The reform also targeted gender inequality in self-determination by supporting teenage girls to make their own decisions on sexual behavior.

To examine the causal effects of sex education for individual outcomes throughout the life cycle, we obtained various unique and exceptionally rich data and applied state-of-the-art estimators of the difference-in-differences (DiD) approach with differential treatment timing (Callaway and Sant'Anna 2021; Borusyak, Jaravel, and Spiess 2024). The share of negative weights totaled one fourth in the two-way fixed-effects model, making our case representative of many other studies with multiple treatment timings where linear regression leads to biases in the reform estimates.

Sex education became an obligatory element of the Swedish primary school curriculum in 1942, before its inclusion became legally binding in 1955. Although mandated by authorities, its practical implementation was supported by national and regional authorities between 1942 and 1958 through the organization of local courses for primary school teachers. Such a gradual implementation of the courses provides us with the municipality's entry to sex education treatment, which we utilized in the identification of the causal effects: Dependent on where primary-school-aged students lived at the time of the reform (cohorts born 1930–1946), they either received sex education or had never done so. We used rich register data on the full population of cohorts under study and their children in a variety of life-cycle outcomes from 1950 until 2017.

We find that the reform yielded significant effects for most intended outcomes for both sexes. First, the reform led to a significant reduction in sexual activity and unwanted pregnancies, as evidenced by a decreased likelihood of abortion and cervical cancer (for women), and a decline in marriages of convenience related to out-of-wedlock births (for both men and women). The reform delayed marriage and childbearing, particularly beyond the age of 25 years, and positively impacted educational outcomes. Second, the reform significantly increased individuals' earnings by 9.3% to 10.9%, through either direct effects in the form of high returns on non-cognitive skills fostered by the reform (preferences and responsibility) or indirect effects via the postponement of marriage and childbearing. Third, it significantly increased women's labor force participation by 1.5 percentage points (or 4.8% of pre-mean) at the ages of 30–35 years and decreased the probability of women to be employed as maids.

Many studies have been conducted on the effects of sex education, but they have primarily discussed teenage health and pregnancy outcomes. According to a recent review of randomized control trials, interventions that simultaneously provide preventive education and contraceptives significantly lower the risk of unintended pregnancies (Oringanje et al. 2016). However, observational studies that applied causal strategies have provided mixed evidence (Kearney and Levine 2012; Sabia 2006).

Programs that propagate abstinence yield no desired effects (Paton, Bullivant, and Soto 2020; Carr and Packham 2017). All these studies, however, have assessed the effects of sex education in modern contexts, where students' preferences in sex activity have been shaped by various informational channels. We contribute to the literature by considering a historical sex education reform in a context in which effective contraception or alternative information channels, like media, were absent and by looking beyond adolescents' outcomes.

Our study makes several contributions to the broader economic literature. First, we explore a school-based sex education reform intended to build and transmit new social norms of family planning and individual and social responsibility among both men and women. The growing body of research on the role of social norms and identities (Guiso, Sapienza, and Zingales 2011; Akerlof and Kranton 2010), and gender identities in particular (Bertrand et al. 2021; Goldin 2014), emphasizes how these norms lead to specific educational and occupational choices. Previous quasi-experimental studies have found large impacts of different shocks to norms on the economic outcomes of women (e.g., the introduction of contraceptive pills, in Bailey and Lindo 2018; Guiliano 2018).

In addition, since our study focuses on sex education that aimed to affect individual and social responsibility, our findings add to the literature on the economic returns to investment in children's non-cognitive skills (Heckman and Mosso 2014). Literature based on randomized educational experiments has demonstrated the possibility of altering students' preferences that are linked to future economic success and differ between genders, such as risk attitudes or competitiveness (Dhar, Jain, and Jayachandran 2022; Kautz, Jagelka, and Heckman 2019; Shurchkov and Eckel 2018). Several observational studies have assessed how school curriculum affects students' political attitudes (Cantoni et al. 2017; Clots-Figueras and Masella 2013).

Our findings provide direct policy implications, despite the obvious difference of the historical Swedish context studied from today. We draw attention to the fact that the policy effects of sex education are long-lasting because they embed in today's population through different channels, including health behavior, family formation, and labor force performance. To the ongoing debates on sex education in both Europe and the US (UNESCO 2018), we contribute evidence of the long-term potential of school-based sex education that includes both preventive and normative components.

## 2 Context

Sweden entered the 1930s with an economic and demographic crisis. The recession hit working families hard, leaving two-thirds in poverty (Schön 2017). The country also faced simultaneous population issues: birth rates among married couples fell faster than in any other European country, the age at first intercourse fell, and the share of children born out-of-wedlock rose (Lundberg and Åmark 2001). Society was traditional, characterized by a large divide in productive activities: men occupied almost all industrial jobs and women worked in domestic spheres but commonly left their jobs after the birth of the first child (Stanfors and Goldscheider 2017). The government's solutions to these societal problems were the prohibition of contraceptives and criminalization of abortion. However, such measures proved inefficient in the worsening socio-economic conditions; when faced with a choice between poverty with children or a substantially better living standard without them, most young couples opted not to have children (SOU 1936).

The rise of the Social Democratic Party and its involvement in various governments gave momentum to efforts to solve the population crisis through welfare reforms and radically remodel ideals of the individual and the state. Figure 1 illustrates the growth of public support for social democrats and the emergence of the social norm governing sexual life, as measured by the number of newspaper articles on the related topic that was taboo before the 1930s. By the advocacy of Gunnar Myrdal, then head of Sweden's population commission, and Alva Myrdal, a social democratic politician, the party's ideological solution was to implement a program that combined social reform and full sexual enlightenment, and thereby shape new societal identities "with aptitudes for personal independence and for collective cooperation" (A. Myrdal and G. Myrdal 1997 [1934]). The Myrdals emphasized the importance of sex education for young individuals, arguing that targeted education, not the banning of abortions or contraceptives, would better impact teenage sexual activity and build a positive view on the family.

[Figure 1 about here]

As a result of these political efforts, the sex education reform became embedded in the general development of the Swedish welfare state, which intended to instill better hygienic and moral standards in society and give everyone a positive attitude toward the family. In the late 1930s, the Swedish Parliament approved a new abortion law (which allowed abortion on medical and social grounds), a new employees' protection law (which prohibited the dismissal of women because of marriage, pregnancy, or childbearing and provided a 12-week parental leave), and in-kind transfers for childbirth and childbearing (which were intended to help women "become dependent on the state instead of husbands") (SOU 1954). Soon the Swedish Association for Sexual Education organized courses for adult sexual literacy, but they never proved popular; nevertheless, the expansion of sexual education was stipulated in the early 1940s, after the introduction of a comprehensive sex education in primary school.

## 3 The 169 Royal Decree on Sex Education

The implementation of sex education among primary school children was easier compared to that for older children, due to the characteristics of Sweden's comprehensive primary education system in the early 1940s. Primary education was both free of charge and compulsory, requiring students to attend school until they completed the highest grade available in their school district's primary school (*Folkskola*). There were a total of 2,487 school districts, which corresponded to municipalities as administrative units, with an expection of 18 small municipalities that formed school districts with neighboring municipalities (Kungliga Skolöverstyrelsen 1942). Children in Sweden typically started school at the age of seven and finished by the age of 14 or 15, which provided an extended period for teaching sex education. Less than 5% of children continued education in secondary schools (Ekstedt 1976).

On April 10, 1942, the Swedish Parliament issued royal decree No 169 that mandated the implementation of sex education in primary schools. Prior to the decree, Swedish authorities had introduced biology of sex and race in the girls' upper secondary school but reconsidered such a restricted focus in favor of socio-democratic values after WWII erupted. Before the decree, the primary school curriculum did not include topics addressing norms and values, except for the historical narratives in the Religious Knowledge subject where students learned the history of Christianity, read the Bible, and reflected on the values of Christian faith. Sex education became compulsory in Sweden in 1955, with full implementation arriving in 1959. This made Sweden the first country in the world to introduce comprehensive sex education in schools (Zimmerman 2015).

With the introduction of sex education, primary schools became the venue for providing students with information about sex matters. As opposed to parents, schools could cover every student and provide uniform and clear information. With this new role, through sex education classes, teachers could translate information on desirable social norms: "Most young people have certainly got to know in their own homes what their parents and society expect from them; it is the school's responsibility to see that it is made clear to all, without exception, what this is" (Royal Board of Education 1964, p.12).

The 1942 decree ordered a change in the curriculum of *Folkskola* in several subjects, where the course content differed by the student's age (Royal Board of Education in Sweden, 1945). Following the decree the authorities published instructions for teaching sex education in school—the guidelines in sex education in 1942 (Ecklesiastiskdepartementet 1944) and an instruction booklet as a separate book in 1945 (Royal Board of Education in Sweden 1945). Given the novelty of the subject, the booklet provided a detailed account of teaching methods, including sample lectures and answers to pupils' potential questions. Below we provide the most essential elements of these instructions and refer the reader to Appendix A for additional details.

One of the common messages was abstinence. Because of the high and previously unmet demand for sex education among students of teen ages, schools sought to provide information on sex matters but by no means encouraged teens to have sex. In junior ages (7 to 10 years old), the class taught students about biology of sexes, but in the grades four to six in elementary school (ages 11–13 years), sex education classes taught about preventive methods. The most common message was abstinence from sexual activity among students but with a broader understanding of the practice. The instructions desexualized masturbation (by destroying the myth that this activity may lead to disease and describing it as something that both boys and girls could perform) and encouraged abstinence to prevent potential life crises rooted in sexual activity among teenagers (e.g., cases of sexual abuse or homosexuality). Abstinence was touted as a “source of power that has a developing effect on the personality and on all aspects of intellectual and physical activity” (SOU 1946).

Although sex education instruction intended to prevent youth pregnancies, it did not intend to alter the desired number of children. The reform has to be understood in light of the ongoing ‘population crisis’ in Sweden, where policies aimed at increasing overall fertility while keeping teenage fertility low. For students aged 14–15 years, instruction posited intercourse as a cause of birth, which was “a rich source of happiness” (p.50). It also paid attention to students’ future role of builders of families, with emphasis on responsibility as parents and the questions of late marriage and marriage-partner choice: “The communion of two people is not a private but a community affair” (p.83). The need for abstinence was largely motivated by the intention to inform of the risk of unintended pregnancies in adolescence and outside marriage, not moral standards.

Another common message in the instructions was responsibility. For students aged 14–15, sex education intended to teach about individual and social responsibility, in line with the idea that public education could solve social issues and raise responsible citizens (Lindgren and Backman Prytz 2021). One such issue was to foster understanding that it is the students’ parents, and often society, too, who bear the economic and other consequences of the students’ early engagement in sexual relations. Another issue was to teach students about their interaction with society. Following the Myrdals’ intention that sex education must form responsible citizens, sex education provided knowledge on the laws and institutions connected to sexual life, such as sterilization, legal and illegal abortion, prosocial behavior, and welfare support for families (e.g., child allowances, home loans, and job security). Teaching on the laws and institutions regulating citizens’ sexual life was revolutionary, deemed to exemplify the functioning of a democratic society.

Finally, sex instruction in school encouraged students to think about the ways they would talk to their own prospective children on sex matters, ideally delivering the messages that they learned as students in school and through cooperation with a school medical officer (Royal Board of Education 1964). Sex instruction intended to diminish the influence of parents on students’ choices in sexual behavior. Students were recommended to read on the current scientific and societal knowledge about sex life when they become parents themselves and talk to their own children.

## 4 Teacher Training Courses

The Royal Board of Education (*Kungliga Skolöverstyrelsen*) did not assume that sex education could be introduced in all schools immediately. The teachers had not received any training in the matter themselves and might have potentially been resistant to teach sex education lessons. To ensure and support the introduction of the new curriculum and provide training to teachers, the Royal Board of Education organized teacher training courses throughout the country.

The teacher training courses were organized in different locations across Sweden. Potential locations were numerous: Sweden was divided into 52 school inspection areas, each of which included between five and 120 school districts (2,570 districts in total). Before organizing the course in a particular location, the regional inspectors contacted local authorities regarding the demand for a course and chose a suitable date and location for the course instructors. Beyond collecting information on whether the inspection area had previously received training, we did not find any indication of a systematic rollout where the Board prioritized some inspection areas over others. The course instructors were medical doctors, teaching professionals, and volunteers through the Swedish Association for Sexual Education. Apart from the courses delivered by the Royal Board of Education, regional inspectors and school districts' principals also arranged their own courses.

The training courses were closely tied in content to the 1942 instruction booklet. The program of a one-day course could, for example, entail three 45-minute lectures and three 35-minute demonstration classes for grades 1 (biology of the sexes), 5 (abstinence and unintended pregnancy), and 7 (social norms). Since sex education was introduced in multiple subjects, primary school teachers from all subject areas participated. A survey conducted by the Royal Board of Education in 1954–55 showed that the majority of teachers were satisfied with the course in terms of the content and number of lectures and demonstration classes (Riksarkivet 1943-1958).

We collected data on all teacher courses from several national archives. The archival sources contain information on the planning and organization of all sex education courses throughout Sweden as well as correspondence with regional school inspectors. The information available includes the school year for the respective course, its location (municipality and school district), and frequently, the school districts covered by the course and number of participating teachers. The annual correspondence contained detailed information on the demand for and implementation of the courses in the previous years, hence we were able to verify information and obtain the complete list of courses. Figure A.1 in Appendix A provides the number of courses by the year of implementation between 1942 and 1958, a total of 230. Courses were discontinued from 1959 due to low demand from the regional inspectors.

## 5 Conceptual Framework

The 1942 sex education reform is a prime example of how schools can transmit social norms to the minds of young individuals. This reform could directly affect individual economic well-being in two distinct ways: (1) The reform explicitly provided new knowledge to students and educated students about social norms and expectations regarding sexual activity, thereby altering students' preferences. (2) The reform provided input to the development of students' personality skills, such as personal responsibility and internal loci of control—both important components of child human capital that improve future earnings in adulthood. It encouraged students to become forward-looking and responsible, and, particularly girls, to exercise the right to protection from sexual abuse; therefore, the reform affected the individual's choice of education, entry into the labor market, and occupation.

First, we propose that the sex education reform translated social norms to individual preferences, particularly regarding sexual activity. Sex education provided new information to students and signified social norms, which enter the individual's utility function and drive behavior (Akerlof and Kranton 2000; Oettinger 1999).

It is well-documented that the introduction of sex education aimed at providing both knowledge and value judgments (norms) (SOU 1936). The reform tackled a previously taboo subject on sexual activity and family formation and imparted previously undisclosed information to students. If sex education provides new knowledge to students on the long-term consequences of sexual activity, it will influence the decision to be sexually active by altering the expected costs or benefits.

Sex education reduces the perceived utility of sexual activity and is expected to ultimately decrease the number of sexually active individuals and unintended pregnancies, (Oettinger 1999) (see Figure B.1 Appendix B). For example, such education may highlight the benefits of abstinence or the individual and societal costs associated with sexual activity. These components were integral to the sex education reforms implemented in Sweden between 1942 and 1958. Abstinence during the teenage years was promoted as a social norm and a means of personality growth, while unintended pregnancies and venereal diseases were framed as immense costs for students, their families, and society.

At the same time, the curriculum highlighted the value of family life once maturity is reached, reflecting the pro-natalist intention of the reform in the context of the 'population crisis'. Institutions such as schools shape norms and identities (Kranton 2016) and – by being taught in schools – norms enter the individual's function and influence behavior (Akerlof and Kranton 2000). The reform placed significant emphasis on whom the society considers an ideal family member: rational individuals who conceive children when the necessary economic resources to support a family are available (Boethius 1985). A key attribute of the reform was, through sex education, to help young individuals to adapt to ongoing societal change and foster a sense of citizenship and identity (A. Myrdal and G. Myrdal 1997 [1934]). While being introduced in the context of the 'population crises' and pro-natalist ideals, it framed the use of contraceptives inside and outside marriage as legitimate for economic reasons. For girls, sex education translated into an awareness of their own rights regarding sexual activity and the de-stigmatization of working mothers.

A likely second way in which the sex education reform influenced behavior – beyond altering utility functions through information and norms – was through the development of personality skills. Personality skills, along with preferences, are malleable in adolescent ages and they affect lifetime economic well-being. Children's time preferences and personality changes, such as becoming more conscientious, at ages 13–15 years predict early-career entry, eminence, and various health and labor outcomes in prime working ages (Hoff et al. 2021; Golsteyn, Grönqvist, and Lindahl 2014). Through role models, schools have affected students' prosocial inclination (Kosse et al. 2020). Regardless of cognition, such non-cognitive skills as responsibility, grit, self-efficacy, and prosociality are highly valued in the labor market (Kautz, Jagelka, and Heckman 2019). In relation to sex education, evaluations of randomized control trials have shown that sex instruction with abstinence and ethical components improves gender equitable skills, confidence, and self-identity among adolescents shortly after the intervention (N. A. Constantine et al. 2015; Louise A. Rohrbach et al. 2015).

## 6 Empirical Strategy

We intended to obtain average treatment effects of the 1942 sex education reform on treated municipalities. We therefore adopted an identification strategy that relied on the gradual rollout of sex education across Swedish municipalities in 1942–1958. The institutional context of the reform gave us two sources of variation: (1) in cohorts who—never or ever—attended the sex education classes in the fifth to eighth grades of primary school in a particular year between 1942 and 1958; (2) in municipalities that eventually introduced the reform or had never done it. The combination of these two sources of variation formed a DiD approach with staggered adoption and allowed us to exclude the influence of the following biases: time (i.e., cohort) bias, which was important because all outcomes we studied changed across cohorts; and selection bias, arising from the potential self-selection of municipalities into treatment. In the identification phase, we did not assume exogeneity of the reform; as shown below, we instead depart from a set of other, more plausible assumptions.

The most common method to estimate the DiD effects is to run a two-way fixed effects regression on the sample. However, the recent methodological literature has cautioned against using regression when the reform is staggering (Roth et al. 2023). In such a case, the regression makes comparisons that a researcher is unwilling to make—it compares later- with earlier- and always-treated municipalities, creating negative weights and biasing the causal effect when the local effects (i.e., across municipalities or years) are heterogeneous. We anticipated the issue in our estimation sample, given that the proportion of earlier- and always-treated municipalities exceeded 50%. Appendix C presents the results of the pre-tests for the presence of negative weights and heterogeneous effects (Roth et al. 2023). We discovered that the effects were indeed heterogenous, and a quarter of the estimation weights for the ever-treated group were negative. Consequently, the use of the two-way fixed effects estimator with constant effects would provide a DiD estimate that would significantly deviate from the true treatment effect.

We therefore implemented a DiD approach by means of the group-time estimator developed by Callaway and Sant'Anna (2021). The group-time estimator is one of the potential (and numerically similar) solutions for the case of staggered reform adoptions when heterogeneous effects are present; this estimator is more attractive owing to its weaker assumptions and more conservative inference (for discussion see Roth et al. 2023).

The estimator's building block is the group-time average treatment effect on the treated, $ATT(g,t) = E[Y_{i,t}(g) - Y_{i,t}(\infty) \mid G_i = g]$, which gives the average treatment effect at time $t$ for the cohort first treated in time $g$. The estimator defines time as a single cohort (i.e., year of birth in our case) and group as the year (first cohort) when the municipality or a group of municipalities received treatment, with never-treated municipalities assigned into a separate category. Under the staggered versions of the parallel trends and no anticipation assumptions, one can identify $ATT(g,t)$ by comparing the expected change in the outcome for cohort $g$ between periods $g - 1$ and t to that for a control group not-yet-treated at period $t$. Clearly, the estimator makes all comparisons relative to the *last pre-treatment period*, which is reasonable as we would like to capture sharp changes in the cohorts' outcomes related to the reform. Not-yet-treated units can include the units treated later, never-treated, or both groups, but earlier-treated municipalities are never included into the control group.

To estimate the ATT effects, the estimator employs a canonical two-way fixed-effects regression and conducts estimations separately for each $g$ and $t$:

$$Y = \alpha_1^{g,t} + \delta_m^{g,t} + \alpha_2^{g,t} \cdot 1\{T = t\} + \beta_3^{g,t} \cdot \left(S_g \times 1\{T = t\}\right) + \varepsilon^{g,t} \quad (1),$$

where $S_g$ denotes a "group" (in our case, the first year when all primary school teachers attended sex education training courses in the municipality of residence) and $t$ is a year of birth. In (1), we control the whole set of municipality fixed effects via $\delta_m^{g,t}$. The coefficient $\beta_3^{g,t}$ is the DiD indicator. In our baseline estimations, we only include never-treated municipalities ($S_g = 0$) in the comparison group because comparing to this group requires a "weak" parallel trend assumption—parallel development only for one year prior to the treatment. However, we recognized that our results would not be sensitive if we additionally included not-yet-treated municipalities or used only them as a comparison (see Section 8.2 for the results).

Subsequently, group-time average treatment effects are aggregated into an overall effect of receiving sex education. First, single ATT($g,t$)'s (i.e., the estimates of $\beta_3^{g,t}$) are aggregated for each group $\tilde{g}$ across all their post-treatment periods $\mathcal{T}$ using the formula: $\theta(\tilde{g}) = \frac{1}{\mathcal{T} - \tilde{g} + 1} \sum_{t=\tilde{g}}^{\mathcal{T}} \beta_3^{\tilde{g},t}$ *(2)*. Then, these effects are averaged together across groups and weighted according to group sizes to form a summary DiD effect, the main aim of the estimation: $\theta^0 = \sum_{g \in G} \theta(g)\, P(G=g \mid G \leq \mathcal{T})$ *(3)*. Therefore, the aggregated effect $\theta^0$ is the average treatment effect of participating in the treatment experienced by all units that ever participated in the treatment and hence resembles the ATT in the canonical DiD setup with two periods and two groups. We obtain more conservative inference by means of a simple multiplier bootstrap procedure (Callaway and Sant'Anna 2021). We also clustered the errors at the municipality of residence level to account for the dependence of groups over time.

The main identification assumptions of the DiD with a group-time estimator are the absence of anticipation, stable unit treatment value, and parallel trends (Callaway and Sant'Anna 2021). We collected detailed data on reform implementation and hence ensured that anticipation—the behavior of the units mimicking the presence of treatment before the actual treatment occurred—was highly unlikely. The stable unit treatment value assumption was also unlikely to be violated in our work because students typically attended the schools in the same municipalities where they resided (Ekstedt 1976). The group-time estimator requires that pre-trends must not exist for one year prior to the reform. To examine the presence of pre-trends and the effects' dynamics, we also estimate the ATT($g,t$) effects aggregated by event-years. The group-time estimator estimates single ATT($g,t$)'s by group and event years: $\theta^e = \sum_{g \in G} 1\{g + e \leq \mathcal{T}\}\, P(G=g \mid G + e \leq \mathcal{T})\mathrm{ATT}(g, g + e)$ *(4)*. For instance, the effect at $e = 0$ occurs for the first cohort affected by the reform. Then the effects are aggregated by group sizes using *(3)*.

However, the parallel trends assumption may be violated even in the event of no visible pre-trends (Rambachan and Roth 2023). In the context of the sex education reform, our treated and control municipalities might have followed distinct development paths even in the absence of the reform, attributable, for instance, to different social norms or labor-market opportunities. As such, we imposed a relaxed, conditional parallel trends assumption that required a parallel trends assumption to hold conditional on covariates and hence added an extra degree of robustness (Roth et al. 2023). To obtain the conditional ATT(*g,t*) values, we used the regression adjustment procedure in conjunction with the group-time estimator (Sant'Anna and Zhao 2020).

Under the conditional parallel trends, this procedure estimates the conditional expectation of the outcome among untreated units and then averages these predictions using the empirical distribution of covariates among treated units:

$$\hat{\tau} = \frac{1}{N_1} \sum_{i:D_i=1} ((Y_{i,t} - Y_{i,g-1}) - \widehat{E}[Y_{i,t} - Y_{i,g-1} \mid D_i=0, X_i] \quad (5),$$

where $\widehat{E}[Y_{i,t} - Y_{i,g-1} \mid D_i=0, X_i]$ is the estimated conditional expectation function fitted on the control units but evaluated at $X_i$ for a treated unit. After estimating the conditional ATT(*g,t*)'s, they are weighted and averaged as in *(2)* and *(3)*. To ensure that the exposure to sex education did not coincide with other relevant local changes during this period of welfare state building, we identified the conditions as a rich set of covariates at the municipality level: urbanization, primary-school expenditure, sex ratio of the birth cohort, share of votes for Social Democratic Party, and the proportion of working women. They are time-varying and measured in the year prior to the reform. Since the estimated propensity score based on these covariates had limited overlap among several small treatment groups in the estimation sample, we used binary indicators of the covariates, divided at their median value.

The group-time estimator that we used in the main body of the paper is a generalization of other approaches proposed in the methodological DiD literature (for discussion see Roth et al. 2023). An alternative is to use an imputation approach proposed by Borusyak, Jaravel, and Spiess (2024), but it requires a researcher to be more confident about parallel trends. In the robustness analysis in Section 8.2, we conducted estimations using the imputation approach and obtained estimates that were nearly identical to our group-time estimates.

## 7 Data

### 7.1 Data on the Rollout of the 169 Royal Decree

Based on our data for sex education training courses for primary schoolteachers, we obtained the coverage by sex education courses for each municipality for each year in 1942–1958. Teachers were not individually responsible for enrolling in the course; instead, it was the responsibility of either the school district, municipality, or a regional principal. Among the school districts that were enrolled in multiple courses across different years, we found very few cases where the proportion of participants differed; instead, municipalities either participated fully or partly (i.e., where the number of participating schol teachers were below the number of all school teachers in the municipality). We then assigned full treatment to the following municipalities: locations where the course was organized, school districts with a 100% participation rate, and neighboring municipalities within a 100-kilometer radius, when the archival data state that the course was also held for the buffer area. As such, we could ascertain that the implementation of the treatment did not begin before the year of implementation assigned based on full treatment (i.e., no anticipation).

As shown in Figure 2, our treatment data revealed significant variation in the entry to treatment over time, with less than half of the municipalities remaining untreated by 1958. The coverage by sex education across cohorts aligns closely to the findings from the national survey on sex habits among Swedish adults conducted in 1967 (see Figure A.2 in Appendix A). The entry into treatment occurred gradually across different waves of courses, starting with 480 unique municipalities covered by early courses organized by school districts or inspectors themselves. Subsequent waves of courses organized by the Royal Board of Education treated additional municipalities, reaching 409 by 1953 and 443 by the end of 1958. By 1959, 1,211 municipalities (47%) remained uncovered by the reform. When accounting for population size, treated municipalities comprised 65%, whereas untreated municipalities constituted 35%.

[Figure 2 about here]

We investigated the factors that explain the earlier initiation of the sex education reform in Appendix D. In the identification phase, we did not assume exogeneity of the reform. However, understanding the driving forces behind the reform at the municipality level support the selection of pre-treatment variables for the conditional DiD estimator. As we found, municipalities that introduced the sex education reform spent more on primary schools and had less conservative social norms (as compared to never-treated municipalities), as exemplified by the lower share of unemployed women and higher share of votes for the Social Democrats in the last national election. Population size does not impact the initiation, suggesting that there is no urban-rural divide (and therefore no differential economic development) between the treatment groups of municipalities. Nevertheless, we use all these covariates in the conditional DiD estimations.

### 7.2 Microdata

We combined the reform data with microdata (Swedish Interdisciplinary Panel, hosted at the Centre for Economic Demography, Lund University), which contain panel information for 1950–2021 linked through unique personal identifiers from various administrative registers for all cohorts born since 1930 as well as for their parents, siblings, and children. Lazuka (2020) previously described such data, their sources, and completion. We selected from these data information on individuals who were aged 11–15 years in 1942–1958—cohorts 1930 to 1946. The estimation sample provided information for 1,026,358 individuals.

The data allowed us to study the outcomes in health outcomes, family formation and career across the full life cycle of the individuals, both in terms of averages and for specific age groups, up to the ages of 70 years for the first generation and 36 years for the second generation. Additionally, we made a richer use of inpatient registers with outcomes such as abortions, cervical cancer, and sexual abuse. We constructed an occupational score based on the occupation and socio-economic index (Lambert et al. 2013). Although we were primarily interested in the effects across the working life cycle, we had to analyze outcomes across different age ranges due to the varying frequencies and starting years of the register data. Sweden's Census of Population from 1950 and 1960 provided an accurate municipality of residence close to the time when the study cohorts were in primary schools.

A group-time estimator requires at least one pre-treatment cohort for each group. The first available cohort in our microdata was composed of individuals born in 1930 (i.e., they were 16 years old in 1946). Thus, our first treatment group was 1946 instead of 1942, and our estimation sample included 1,939 out of the full pool of 2,503 municipalities. This should not be a problem for external validity of our results due to the following reason: Since large and less conservative municipalities implemented the reform earlier, particularly major cities like Stockholm, which organized courses in the early 1940s, we set our final estimation sample to consist of a more homogeneous group of treated and untreated municipalities. These municipalities were mostly covered by the courses organized by the Royal Board of Education.

## 8 Results

### 8.1 Sexual Activity

We start with the DiD estimations for the health outcomes related to sexual activity or abuse in ages 34–40 years, the youngest possible from the inpatient hospital records, and present the aggregated reform estimates in Table 1 and the event studies for the outcomes with significant effects, in Figure 3. We refer the reader to Appendix E for the event studies related to the remaining outcomes. We find that, owing to the sex education reform, women were much less likely to abort a pregnancy, by 1.1 percentage points (or 79% of the pre-mean). For this and other outcomes, we present event studies for 5 event-years before and after the reform; for post-reform event-years, they essentially show the reform effects for children exposed at ages 15 to 11 at the earliest. For abortion, the large reform effect emerges for all event-years in the post-treatment period, with no evidence of pre-trends. At the time, women could receive approval for abortion for medical reasons, including risk to the woman's life, potential transmission of a serious disease, sex abuse, or a "weakness," which referred to worn-out mothers who already had children (Sjövall 1972). Illegal abortion was punishable with imprisonment, but women were rarely sentenced. Our findings therefore suggest that the reform encouraged women to avoid unwanted pregnancies.

[Table 1 and Figure 3 about here]

Furthermore, we observe a large negative impact on the probability of cervical cancer and inflammatory diseases of female pelvic organs, amounting to 1 percentage points, or 23.5% of the pre-mean. Since cervical cancer is caused by sexually transmitted diseases (Venkatas and Singh 2020), the effect can be attributed to reduced sexual activity, a smaller number of sex partners, or increased use of protection. In contrast to abortions in middle ages, which reflect sexual activity in middle ages, cervical cancer is a measure of sexual activity among teenagers and young adults. Infections with the human papillomavirus (HPV) occur during sexual activities at young ages and cause cervical cancer often decades later. For the variables measured for both men and women, we also report difference-in-difference-in-differences effects (DiDiD) that should be interpreted as a difference in the reform effects for women versus men. We observe no reform effects for sexually transmitted diseases themselves or sexual abuse for the whole sample or for women versus men. Notably, detecting effects for veneral diseases could be challenging for these causes owing to the significantly underrepresented rates of related inpatient admissions, unlike those for cervical cancer that could be treated only within a hospital system and hence were registered in full.

### 8.2 Family formation

The sex education reform was designed to influence students' preferences regarding teenage childbearing and adult family formation. Table 2 presents results on how the sex education reform influenced individual decisions to marry across the life cycle. We first analyze the reform impact on marriage of convenience, defined as marriages occurring within a year after childbirth. We treat marriage of convenience as a direct measure of unplanned pregnancies, unlike other nuptiality and fertility outcomes, which may be impacted by, for example, prolonged education. The results show that the reform significantly reduced the probability of marriage of convenience by age 18, which was the legal age for women to marry (0.17 percentage points or 28% of the pre-mean). As shown with DiDiD effects, this reducing effect is twice the average among women, aligning with the expectation that women, compared to men, bore a larger share of the costs associated with early engagement in sexual activity. We also note a large and statistically significant reform effect for marriage of convenience for the whole life cycle of men and women (-0.4 percentage points or 11.3% of the pre-mean). Therefore, the reform reduced unplanned births outside marriage, as intended.

[Table 2 about here]

Turning to traditional marriages, we do not find any reform effects on the probability of marrying by age 18. Young marriages were very rare in our context and usually occurred in coincidence with marriages of convenience. However, the reform effects on the probability of marrying emerge by age 25, by which time the longest schooling available had been completed and individuals typically embarked on a career. As such, the reform reduced the probability of marriage by 2 percentage points or 4% of the pre-mean. Such reform effects diminished over the life cycle but remained present by the age of 45 years, meaning that the reform delayed marriage and somewhat influenced the proportion of people who never married. While deferral of marriage was not encouraged by the sex education reform, responsibility in the decision of who to marry, was. Postponement of marriage may as well be an indirect effect through prolonged education.

Table 3 presents the results for parenthood outcomes. We find no effects for teenage pregnancies, although we noted that the reform reduced the probability of having a child at the ages of 18–25 years by 1.4 percentage points or 2.8% of the pre-mean. The effect decreased slightly under conditional parallel trend assumptions but remained statistically significant. This finding must be understood in the context of the Swedish population crisis and light of the actual reform. In contrast to more recent sex education reforms, which aim at reducing teenage fertility, the introduction of sex education in Swedish primary schools occurred in the context of low fertility rates and attempts to raise those. Teenage pregnancies were not an issue that needed to be dealt with (pre-mean 3.4%) but could have become a consequence of the pronatalist era, in absence of sex education. The median age of childbearing was 30 years, and for later ages, the reform instead encouraged individuals to increase the likelihood of having a child, by 1.8 percentage points (4% of the pre-mean). A positive effect on later-life fertility is in line with the objectives of the reform, as sex education was introduced as a response to Sweden's population crisis of plummeting birth rates. However, this effect was weak and unclear in the event studies. The reform effects for parenthood did not consistently differ by sex, although they implied that women exhibited larger postponement and catching-up effects compared with men. In sum, across the full life cycle, the reform encouraged individuals to postpone parenthood but did not alter the desired number of children.

[Table 3 about here]

### 8.3 Educational Attainment and Occupational Sorting

Sex education can impact educational attainment and later-life outcomes in two ways: First, by encouraging postponement of childbearing, it allows women and men to stay longer in educational institutions. The sex education reform increased the perceived opportunity costs of childbirth by teaching social and economic sanctions of not following social norms (Royal Board of Education 1964, p. 11–12). The reform by these means encouraged individuals to postpone childbearing, and pursuing higher education can thereby be an indirect effect of the reform. Second, in line with comprehensive sex education, the reform impacted personality skills and internal loci of control, thereby going beyond promoting abstinence. The formation of non-cognitive skills can thus have direct effects on educational outcomes. Moreover, it might also have led individuals to choose specific educational fields associated with occupations involving public responsibilities.

The results for educational outcomes are presented in Table 4 (aggregated DiD estimates) and Figure 4 (event studies). The effects on years of schooling are positive and statistically significant at the 1% level, albeit relatively small (0.04 years or 0.5% of the pre-mean). However, the positive effects on schooling appear much stronger and more reassuring (according to event studies) for the completion of secondary education and vocational training, each increasing by 1 percentage point or 5% of the pre-mean. Ultimately, 20% of individuals obtained such schooling. Thus, the reform improved the likelihood of students successfully completing primary school, given that enrollment in secondary school requires passing a national test and completing all primary school subjects (Skolöverstyrelsen 1955). The same holds true for vocational training, which is provided after the age of 18 years and involves an entrance exam (SOU 1974). The event studies show that the effects on education increase the earlier children enroll in sex education.

[Table 4 and Figure 4 about here]

Regarding the educational fields, we find that the reform improved the probability of individuals completing specific fields of upper secondary and college education. Notably, the implementation of sex education in primary schools stimulated individuals to pursue further education and opt for degrees in social sciences, law, and administration (1 percentage point increase or 10% of the pre-mean). Compared with other fields, the social sciences required a willingness to work for the common good, demanded high skills, and offered employment guarantees amid the growth of public services in this era. Additional analyses reveal that men chose to pursue careers in health care, whereas women opted for teaching professions. This result suggests that for women who grew up in an era where their mothers were not working, future employability played a great role when choosing an educational field, similar to what happened to the US cohorts (Goldin 2021).

Table 5 presents the results for the reform effects for occupational outcomes. Consistent with the results for educational fields, the reform increased the likelihood of being employed in public services (0.8 percentage points or 3% of the pre-mean) rather than in industry or private services. These shifts are driven by the rise in number of not only specialist jobs in health care and teaching but also positions in government and administration. The reform effects on occupations did not only emerge for a narrow group of public services—we find a statistically significant increase in overall occupational score (by 0.3 units, 0.5% of the pre-mean). The entire distribution of the occupational score shifted to the right, with somewhat larger effects at the bottom decile (for women) and top decile (for men and women).

[Table 5 about here]

### 8.4 Labor Market Performance

We expected to find positive reform effects on individuals' permanent income, either through improved non-cognitive skills or as a result of postponed marriage and childbearing and prolonged education. Our results from Table 6 show that the sex education reform increased individuals' permanent earnings income by 9.3% and 10.9% under the unconditional and conditional assumptions, respectively. The magnitude of the effect is striking, given that the reform encouraged individuals to choose jobs in the public sector, which, at the time, did not guarantee high wages compared with employment in manufacturing or the private sector (Schön 2017). The effect on earnings is most pronounced between ages 48 and 50 years, as earnings for both men and women peaked. As DiDiD effects show, at this stage, the effect was larger for women—10.7% versus 6.9% under conditional terms, although the difference is not statistically significant.

[Table 6 about here]

The reform should have impacted women's labor-market participation by postponing childbirth and facilitating personal responsibility and the acquisition of education that enhanced employability. As shown in Table 6, we find that the reform allowed women to enter the labor force (from being out of the labor force to receiving employment) at the ages of 30–35 years (1.5 percentage points or 4.8% of the pre-mean). Such effects decrease in size for the later reproductive ages, consistent with the phenomenon of women getting married and having more children during these stages. However, employment effects are also present for married women, as women also became less likely to be employed in the household sector (as maids or in the small businesses of their husbands), equivalent to an increase in the probability of having an independent occupation (1.5 percentage points or 15.4% of the pre-mean). We do not find any similar effects for men's labor-market outcomes. The reform-driven shift in the employment of women from the unpaid household sector to the paid work and service sector has created significant productivity gains, given that working in the public sector was valued at three to four times as much (Krantz 1987). Therefore, the sex education reform contributed to the acceleration of women's entry to paid jobs—a development that distinguishes Sweden from other developed countries in which this trend was staggering.

### 8.5 Robustness Analysis: The Validity of the Estimator and Interaction with Other Social Reforms

Methodologically, we ensured that the assumptions of the DiD approach were likely to be met. We carefully checked the treatment data for each municipality to avoid anticipation. The event-study analysis revealed no significant pre-trends, and the results obtained were robust to controls for the pre-trends based on various observable regional characteristics measuring the development of the economy and of social norms. However, the DiD approach does not control for the biases stemming from the pre-trends in unobservable characteristics and from time-varying effects, such as the influence of overlapping reforms. We thus assess these effects as well.

We first examined whether pre-treatment trends in unobserved municipality characteristics affected the results. To do this, we ran the estimations, using the group of not-yet-treated municipalities instead of never-treated municipalities as a control.

Municipalities that eventually introduced the reform likely shared many observed and unobserved characteristics. As shown in Figure F.1–F.7 of Appendix F, the results for all outcomes are almost identical to what we report in the main body of the paper.

The methodological DiD literature has proposed several approaches to estimate ATT effects for the case of staggered treatment timing that are robust to heterogeneity over time and space. However, Callaway and Sant'Anna's group-time estimator is a generalization of two other popular approaches. The estimator by Sun and Abraham (2021) produces the same aggregated and weighted ATT, using never-treated or last-to-be-treated units as controls. The same is true for Chaisemartin and D'Haultfœuille (2020)'s estimator, with the only difference being a particular choice of weights. Instead of the group-time estimator, we consider the imputation estimator proposed by Borusyak, Jaravel, and Spiess (2024), which offers a valid alternative. In Table F.8 Appendix F, we perform estimations for women's sexual activity outcomes based on the imputation approach and obtain the estimates nearly identical to the group-time estimates.

We further assess the impact of overlapping reforms. The sex education reform was planned and implemented together with a set of social reforms intended to reduce the costs of childbearing (SOU 1954). Most reforms (e.g., the employees' protection law and childbearing cash transfers) were introduced abruptly across the country and consequently affected all women (i.e., mothers of the first generation). Hence, the effects of such reforms had been absorbed by the birth cohort dummies. One reform—opening of maternity wards with a 10-day laying-in period—occurred gradually across the cohorts and municipalities and had lasting economic effects for the cohorts in their adulthood (Lazuka 2023). We thus study the impact of new wards' openings on our reform estimates in Table F.9 and Figure F.1 in Appendix F and find no interaction effects. Finally, starting in the 1940s, the authorities gradually extended mandatory schooling to 9 years (Holmlund 2020), but our analysis revealed that only 3.9% of students within our cohorts were affected by the schooling reform.

## 9 Conclusions

In the 1940s–1950s, during Sweden's pivotal period of welfare state construction based on liberal values, the government introduced sex education in primary schools, making Sweden a global pioneer for mandatory sex education. The Swedish sex education reform promoted abstinence and rational decision-making in sexual and family-formation matters but was not limited to these principles. It also served as a pioneering social intervention, emphasizing the notion of responsible citizenship. At today's point, the majority of the students who were affected by the reform have completed their working careers, and their children have entered active working age. We assessed the causal effects of sex education for a variety of outcomes of these individuals throughout the life cycle, based on a state-of-the-art DiD estimators applied to exceptionally rich microdata. We estimated the reform effects under both unconditional and conditional parallel trends assumption and tested for the potential influence of other social reforms.

As our first major finding, we confirmed that the reform led to a significant reduction in sexual activity and unwanted pregnancies. This was evidenced by the decreased likelihood of abortion and cervical cancer (for women) and decline in marriages of convenience related to out-of-wedlock births (for both men and women), with the effects emerging already by age 18 years and lasting for all reproductive ages. The reform also delayed marriage and childbearing, particularly beyond the age of 25 years, but did not alter the desired number of children born. We therefore confirm that the school may permanently affect preferences in sexual activity and family formation (Oettinger 1999).

Second, we found that the reform significantly increased individuals' permanent earnings by 9.3% to 10.9%, which may be a result of changing demographic behavior but also underscores to the importance of responsibility as a key economic trait and its malleability during adolescence (cf. Kautz, Jagelka, and Heckman 2019; Heckman and Mosso 2014). Along these lines, our results also suggested that the reform encouraged socially responsible individuals, as supported by the effects observed for women who were more likely to enter the labor force and for both men and women who became more likely to choose educational fields and professions related to the welfare state expansion.

Overall, our findings support the notion that norms and behavior can be influenced by school curricula (Oettinger 1999; see also Cantoni et al. 2017; Akerlof and Kranton 2010; Kranton 2016). Such effects shape the nation—the life of students across the entire life cycle whose relative share is sizable to alter social norms in society. Although the scope of comprehensive sex education varies across different countries' curricula, most nations include information on sexual activity and its prevention, relationships, as well as discussions on gender, social norms, and life skills. What sets Sweden apart in the introduction of sex education is its early implementation, which provided us with an opportunity to study the effects of sex education that other countries can anticipate. Further study on the longer-term effects of comprehensive sex education in other settings can make valuable contributions to the ongoing debates regarding sex education in school in the US and other countries in Europe.

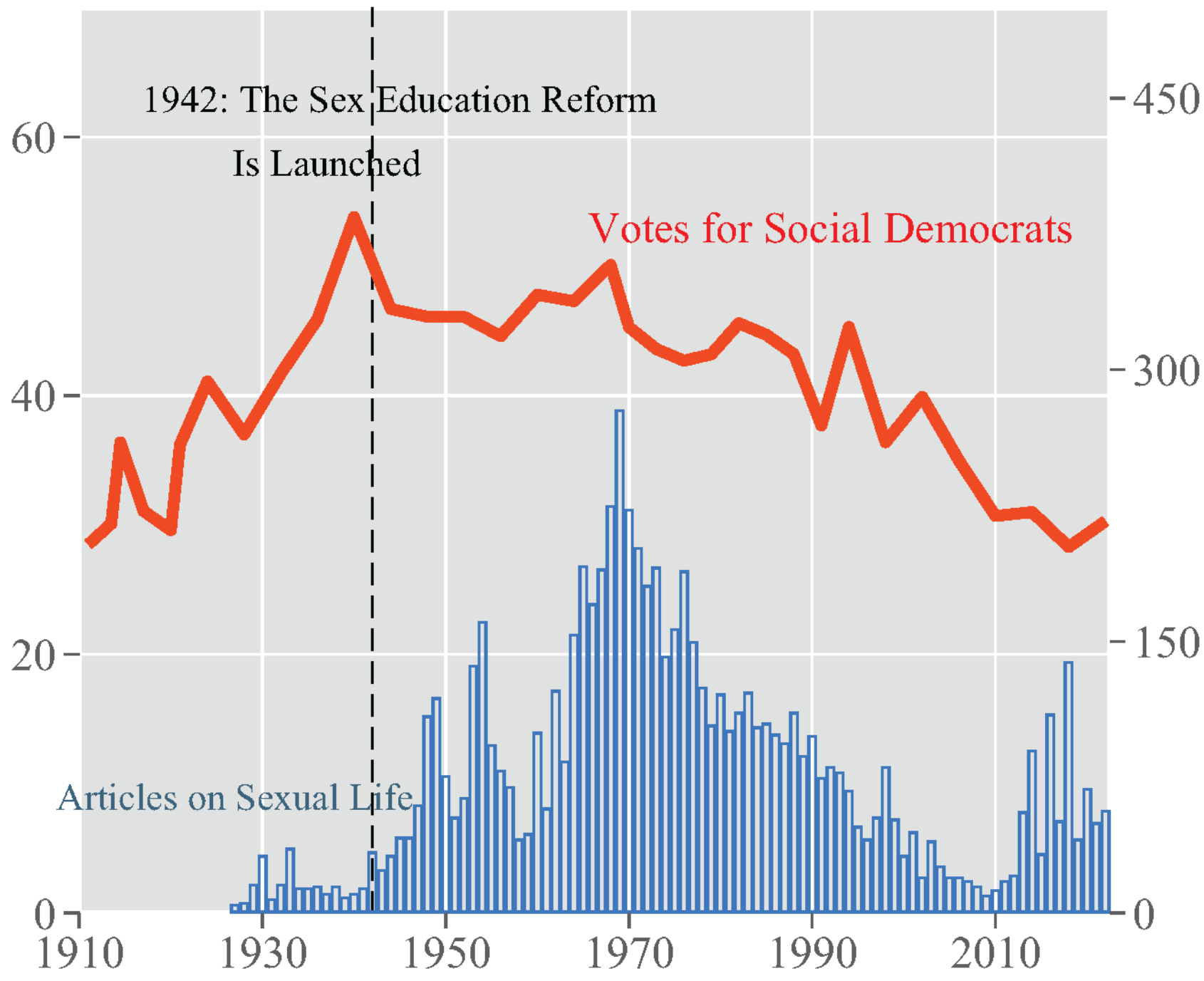

Figure 1 – Social democrats' dominance in the share of votes (left axis, %) and the newspaper articles on sexual life (right axis, totals) during the sex education reform.

*Note:* The share of votes for social democrats is sourced from Statistics Sweden (2023). The number of newspaper articles on sexual life is derived from data containing all newspapers from 1900 to 2020 in Sweden, as listed in the Royal Library (2023).

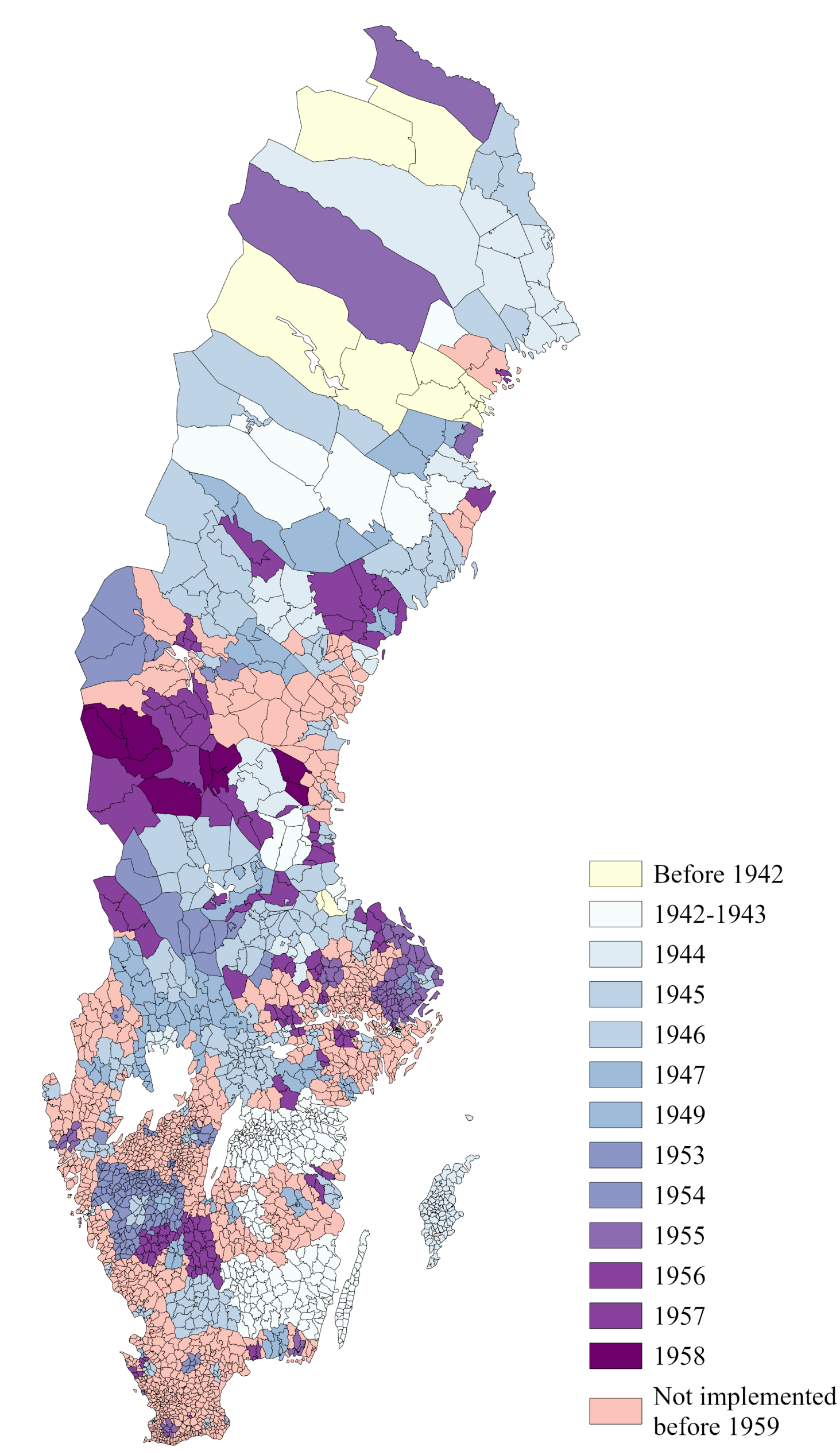

Figure 2 – Staggered entry to the treatment under the 169 royal decree on sex education across municipalities in Sweden.

*Source*: Based on data on reform implementation gathered from the national archives.

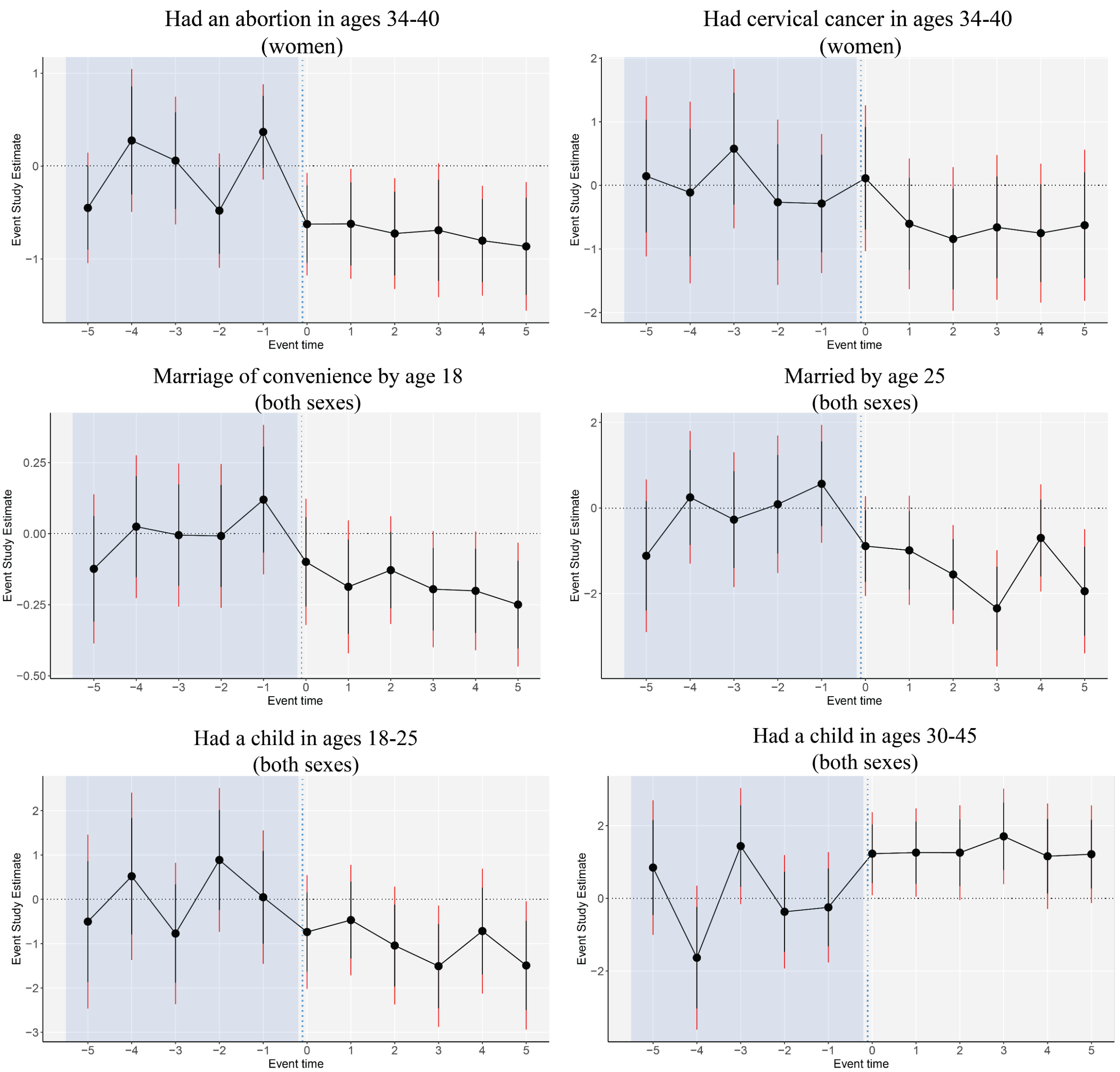

Figure 3 – Event studies for sexual activity and family formation, ages 18–45 years.

*Note*: The figure reports ATT(g,t) effects parameters aggregated by event time and weighted by group size, under the unconditional parallel trends assumption and with clustering at the municipality level. To ease interpretation, we multiplied binary outcomes by 100. The inference procedures used both asymptotic and bootstrapped standard errors (with wider confidence intervals). Both types of standard errors adjust for clustering at the municipality level, while bootstrapped standard errors also adjust for multiple hypothesis testing.

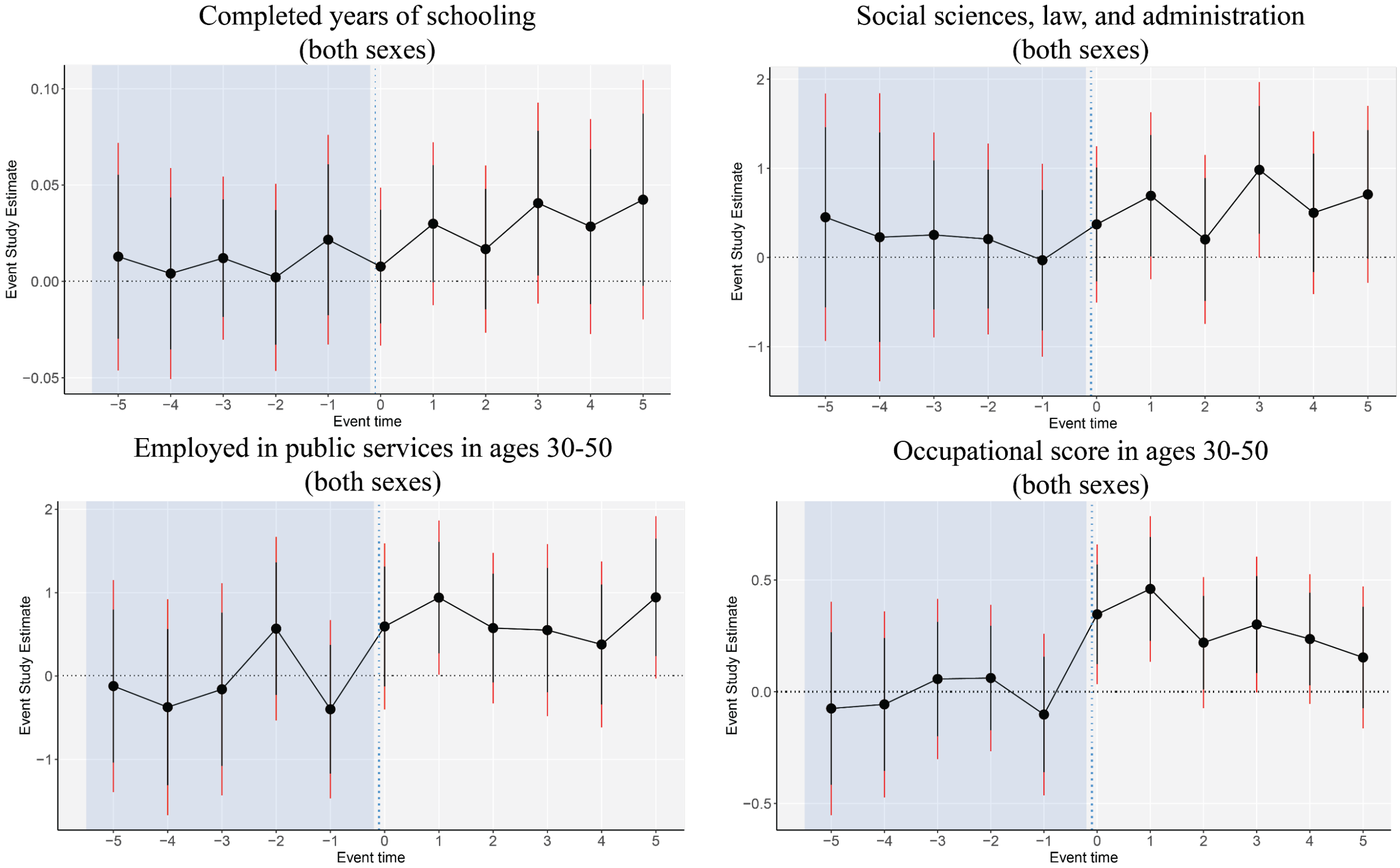

Figure 4 – Event studies for educational and occupational outcomes.

*Note*: The figure reports ATT(g,t) effects parameters aggregated by event time and weighted by group size, under the unconditional parallel trends assumption and with clustering at the municipality level. To ease interpretation, we multiplied binary outcomes by 100. The inference procedures used both asymptotic and bootstrapped standard errors (with wider confidence intervals). Both types of standard errors adjust for clustering at the municipality level, while bootstrapped standard errors also adjust for multiple hypothesis testing.

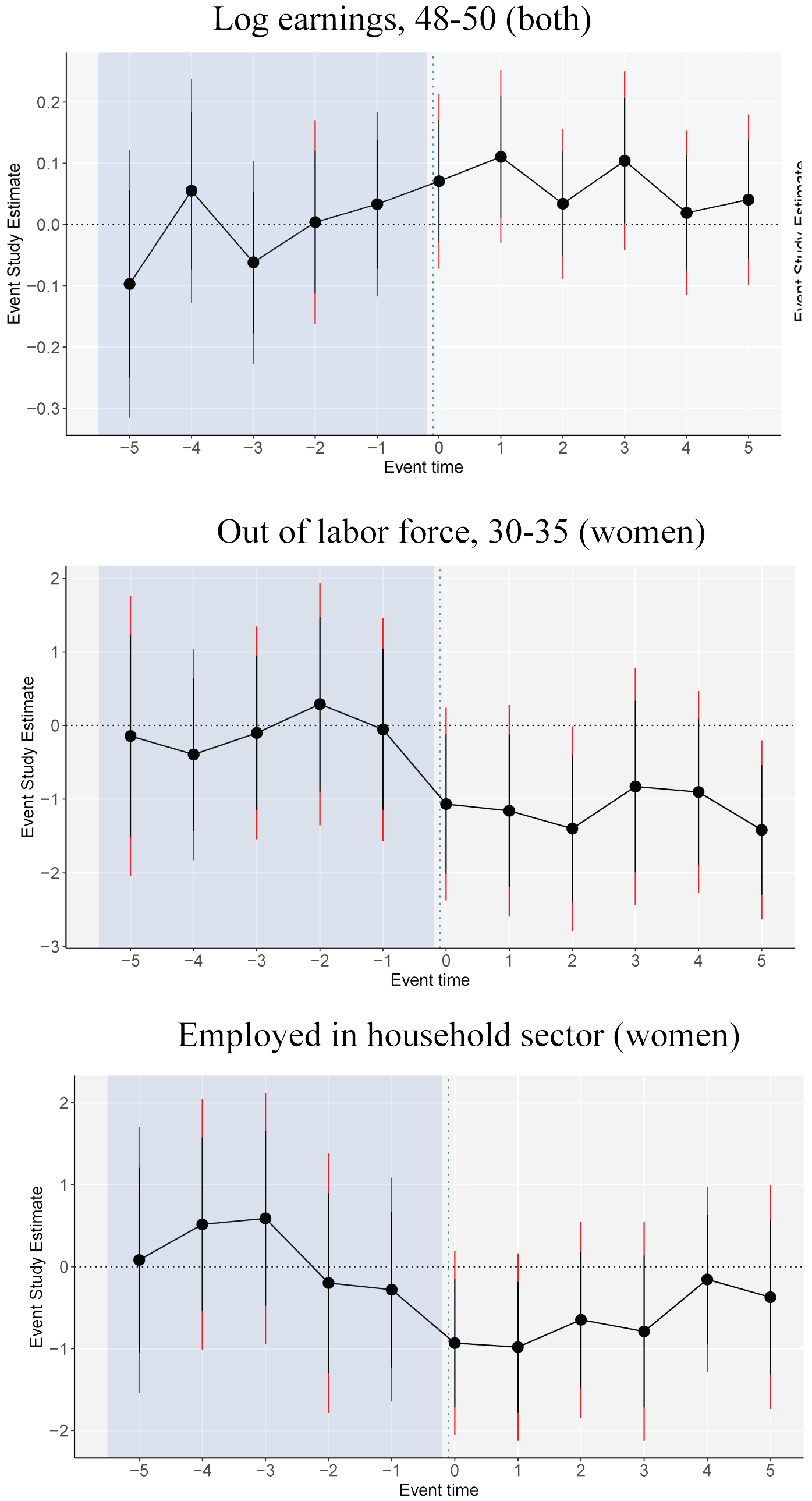

Figure 5 – Event studies for labor market outcomes.

*Note*: The figure reports ATT(g,t) effects parameters aggregated by event time and weighted by group size, under the unconditional parallel trends assumption and with clustering at the municipality level. To ease interpretation, we multiplied binary outcomes by 100. The inference procedures used both asymptotic and bootstrapped standard errors (with wider confidence intervals). Both types of standard errors adjust for clustering at the municipality level, while bootstrapped standard errors also adjust for multiple hypothesis testing.

Table 1 – Effects on sexual activity and abuse, ages 34–40 years.

| | Abortion (women) | Cervical cancer (women) | Sexually transmitted diseases (both sexes) | Sexual abuse (both sexes) |
|---|---|---|---|---|
| **(a) Unconditional DiD** | -1.061*** | -1.024*** | 0.014 | 0.004 |
| | (0.282) | (0.400) | (0.028) | (0.003) |
| **DiDiD, women minus men** | | | -0.014 | -0.009 |
| | | | (0.040) | (0.011) |
| **(b) Conditional DiD** | -1.079*** | -0.986*** | 0.016 | 0.002 |
| | (0.264) | (0.391) | (0.029) | (0.003) |
| **DiDiD, women minus men** | | | -0.004 | -0.013 |
| | | | (0.041) | (0.014) |
| **Pre-mean** | 1.358 | 4.342 | 0.053 | 0.017 |
| **Individuals** | 379,494 | 379,494 | 776,930 | 776,930 |

*Note*: For DiD, the table reports aggregated (average, weighted by group size) ATT(g,t) effects parameters, under the conditional and unconditional parallel trends assumptions, with clustering at the municipality level. For difference-in-difference-in-differences (DiDiD), the table reports the difference between aggregated ATT(g,t) effects parameters (DiD) estimated on the sample of women and men. For presentation purposes, we multiplied binary outcomes by 100. Unconditional group-time estimates were obtained following the estimator of Callaway and Sant'Anna (2021). Conditional group-time estimates in addition apply the regression-outcome DD estimator of Sant'Anna and Zhao (2020) with pre-treatment levels of urbanization, primary-school expenditures, sex ratio of the birth cohort, and share of working women in the municipality as controls. To estimate models at the level of variation, we collapsed our data at the municipality-by-birth cohort cells with weights totaling the number of individuals in each cell. Inference procedures used bootstrapped standard errors adjusted for multiple hypothesis testing and clustering at the municipality level.

*** p<0.01, ** p<0.05, * p<0.1

Table 2 – Effects on marriage, ages 18–45 years.

| | **Marriage of convenience** | | **Married** | | | |
|---|---|---|---|---|---|---|
| | by 18 | 18–45 | by 18 | by 25 | by 30 | by 45 |
| **(a) Unconditional DiD** | -0.172*** | -0.384*** | -0.012 | -2.131*** | -1.709*** | -0.885*** |
| | (0.063) | (0.142) | (0.134) | (0.465) | (0.395) | (0.265) |
| **DiDiD, women minus men** | -0.271*** | -0.053 | 0.077 | 0.707 | 0.476 | 0.317 |
| | (0.100) | (0.225) | (0.196) | (0.610) | (0.526) | (0.397) |
| **(b) Conditional DiD** | -0.169*** | -0.302** | 0.065 | -1.811*** | -1.225*** | -0.631** |
| | (0.064) | (0.150) | (0.135) | (0.499) | (0.409) | (0.266) |
| **DiDiD, women minus men** | -0.249** | -0.101 | 0.213 | 0.168 | -0.544 | -0.075 |
| | (0.104) | (0.244) | (0.199) | (0.672) | (0.581) | (0.433) |
| **Pre-mean** | 0.608 | 3.534 | 2.366 | 54.741 | 77.176 | 86.904 |
| **Individuals** | 1,026,358 | 989,655 | 1,026,358 | 1,020,490 | 1,015,396 | 989,655 |

*Note*: For DiD, the table reports aggregated (average, weighted by group size) ATT(g,t) effects parameters, under the conditional and unconditional parallel trends assumptions, with clustering at the municipality level. For DiDiD, the table reports the difference between aggregated ATT(g,t) effects parameters (DiD) estimated on the sample of women and men. For presentation purposes, we multiplied binary outcomes by 100. Unconditional group-time estimates were obtained following the estimator of Callaway and Sant'Anna (2021). Conditional group-time estimates in addition apply the regression-outcome DiD estimator of Sant'Anna and Zhao (2020) with pre-treatment levels of urbanization, primary-school expenditures, sex ratio of the birth cohort, and share of working women in the municipality as controls. To estimate models at the level of variation, we collapsed our data at the municipality-by-birth cohort cells with weights totaling the number of individuals in each cell. Inference procedures used bootstrapped standard errors adjusted for multiple hypothesis testing and clustering at the municipality level.

*** $p<0.01$, ** $p<0.05$, * $p<0.1$

Table 3 – Effects on parenthood, ages 18–45 years.

| | **Having a child** | | | | **Total number of children** |
|---|---|---|---|---|---|
| | by 18 | 18–25 | 25–30 | 30–45 | |
| **(a) Unconditional DiD** | 0.096 | -1.436*** | -0.420 | 1.771*** | 0.022* |
| | (0.147) | (0.458) | (0.431) | (0.398) | (0.012) |
| **DiDiD, women minus men** | 0.087 | 0.942 | -0.652 | 0.872 | 0.024 |
| | (0.227) | (0.611) | (0.591) | (0.585) | (0.015) |
| **(b) Conditional DiD** | 0.059 | -0.920*** | -0.531 | 1.643*** | 0.025** |
| | (0.140) | (0.440) | (0.451) | (0.420) | (0.012) |
| **DiDiD, women minus men** | 0.093 | 0.357 | -1.199* | 1.361*** | 0.024 |
| | (0.242) | (0.630) | (0.650) | (0.462) | (0.016) |
| **Pre-mean** | 3.462 | 50.188 | 45.277 | 43.242 | 2.008 |
| **Individuals** | 1,026,358 | 1,020,490 | 1,015,396 | 989,655 | 989,655 |

*Note*: For difference-in-differences (DiD), the table reports aggregated (average, weighted by group size) ATT(g,t) effects parameters, under the conditional and unconditional parallel trends assumptions, with clustering at the municipality level. For difference-in-difference-in-differences (DiDiD), the table reports the difference between aggregated ATT(g,t) effects parameters (DiD) estimated on the sample of women and men. For presentation purposes, we multiplied binary outcomes by 100. Unconditional group-time estimates were obtained following the estimator of Callaway and Sant'Anna (2021). Conditional group-time estimates in addition apply the regression-outcome DiD estimator of Sant'Anna and Zhao (2020) with pre-treatment levels of urbanization, primary-school expenditures, sex ratio of the birth cohort, and share of working women in the municipality as controls. To estimate models at the level of variation, we collapsed our data at the municipality-by-birth cohort cells with weights totaling the number of individuals in each cell. Inference procedures used bootstrapped standard errors adjusted for multiple hypothesis testing and clustering at the municipality level. *** $p<0.01$, ** $p<0.05$, * $p<0.1$

Table 4 – Effects on completed education.

| | **Years of schooling** | **Secondary** | **Vocational training** | **College** | **Social sciences, law and administration** | **Natural sciences** |
|---|---|---|---|---|---|---|
| **(a) Unconditional** | 0.043*** | 0.900** | 1.020*** | 0.636 | 0.978*** | 0.138 |
| **DiD** | (0.021) | (0.442) | (0.449) | (0.406) | (0.342) | (0.215) |
| **DiDiD, women** | -0.013 | 0.198 | 0.078 | -0.572 | 0.486 | -0.013 |
| **minus men** | (0.026) | (0.592) | (0.664) | (0.503) | (0.472) | (0.376) |
| **(b) Conditional** | 0.039** | 0.820* | 0.818* | 0.496 | 1.092*** | -0.110 |
| **DiD** | (0.018) | (0.476) | (0.455) | (0.333) | (0.319) | (0.209) |
| **DiDiD, women** | -0.028 | -0.278 | -0.457 | -0.843 | 0.125 | 0.006 |
| **minus men** | (0.027) | (0.658) | (0.743) | (0.516) | (0.514) | (0.376) |
| **Pre-mean** | 7.933 | 19.209 | 19.496 | 8.417 | 9.954 | 4.301 |
| **Individuals** | 969,247 | 969,247 | 969,247 | 969,247 | 969,247 | 969,247 |

*Note*: For DiD, the table reports aggregated (average, weighted by group size) ATT(g,t) effects parameters, under the conditional and unconditional parallel trends assumptions, with clustering at the municipality level. For DiDiD, the table reports the difference between aggregated ATT(g,t) effects parameters (DiD) estimated on the sample of women and men. For presentation purposes, we multiplied binary outcomes by 100. Unconditional group-time estimates were obtained following the estimator of Callaway and Sant'Anna (2021). Conditional group-time estimates in addition apply the regression-outcome DiD estimator of Sant'Anna and Zhao (2020) with pre-treatment levels of urbanization, primary-school expenditures, sex ratio of the birth cohort, and share of working women in the municipality as controls. To estimate models at the level of variation, we collapsed our data at the municipality-by-birth cohort cells with weights totaling the number of individuals in each cell. Inference procedures used bootstrapped standard errors adjusted for multiple hypothesis testing and clustering at the municipality level. *** $p<0.01$, ** $p<0.05$, * $p<0.1$

Table 5 – Effects on sector of employment and occupation, ages 30–50 years.

| | Sector of employment (both sexes) | | | | | Occupational score (both sexes) |
|---|---|---|---|---|---|---|
| | Manu-facturing | Private services | Public services | Teaching, research, healthcare | Government and administration | |
| **(a) Unconditional** | -0.652*** | -0.439* | 0.786*** | 0.491*** | 0.459*** | 0.293*** |
| **DiD** | (0.318) | (0.243) | (0.320) | (0.236) | (0.164) | (0.083) |
| **DiDiD, women** | 1.110** | -0.065 | -0.372 | 0.201 | -0.277 | 0.167 |
| **minus men** | (0.469) | (0.390) | (0.466) | (0.356) | (0.299) | (0.149) |
| **(b) Conditional** | -0.353 | -0.486* | 0.829*** | 0.487* | 0.321* | 0.326*** |
| **DiD** | (0.373) | (0.264) | (0.368) | (0.260) | (0.179) | (0.116) |
| **DiDiD, women** | 0.723 | -0.187 | -0.706 | 0.421 | -0.192 | -0.094 |
| **minus men** | (0.478) | (0.415) | (0.518) | (0.387) | (0.337) | (0.171) |
| **Pre-mean** | 28.271 | 16.269 | 28.879 | 17.105 | 14.579 | 58.145 |
| **Individuals** | 1,010,713 | 1,010,713 | 1,010,713 | 1,010,713 | 1,010,713 | 1,006,639 |

*Note*: For DiD, the table reports aggregated (average, weighted by group size) ATT(g,t) effects parameters, under the conditional and unconditional parallel trends assumptions, with clustering at the municipality level. For DiDiD, the table reports the difference between aggregated ATT(g,t) effects parameters (DiD) estimated on the sample of women and men. For presentation purposes, we multiplied binary outcomes by 100. Unconditional group-time estimates were obtained following the estimator of Callaway and Sant'Anna (2021). Conditional group-time estimates in addition apply the regression-outcome DiD estimator of Sant'Anna and Zhao (2020) with pre-treatment levels of urbanization, primary-school expenditures, sex ratio of the birth cohort, and share of working women in the municipality as controls. To estimate models at the level of variation, we collapsed our data at the municipality-by-birth cohort cells with weights totaling the number of individuals in each cell. Inference procedures used bootstrapped standard errors adjusted for multiple hypothesis testing and clustering at the municipality level. *** $p<0.01$, ** $p<0.05$, * $p<0.1$

Table 6 – Effects on labor market performance, ages 30–55 years.

| | **Log earnings (both sexes)** | **Out of labor force (women)** | | | **In household production (women)** |
|---|---|---|---|---|---|
| | 48–50 | 30–35 | 35–45 | 45–55 | 35–45 |
| **(a) Unconditional DiD** | 0.093*** | -1.436*** | -0.572* | -0.685* | -1.483*** |
| | (0.045) | (0.395) | (0.324) | (0.408) | (0.277) |
| **DiDiD, women** | 0.053 | | | | |
| **minus men** | (0.069) | | | | |
| **(b) Conditional DiD** | 0.124*** | -1.350*** | -0.595 | -0.562 | -1.117*** |
| | (0.044) | (0.406) | (0.368) | (0.403) | (0.218) |
| **DiDiD, women** | 0.044 | | | | |
| **minus men** | (0.071) | | | | |
| **Pre-mean** | 10.684 | 31.881 | 18.349 | 15.903 | 10.318 |
| **Individuals** | 981,934 | 492,074 | 490,729 | 483,308 | 397,757 |

*Note*: For DiD, the table reports aggregated (average, weighted by group size) ATT(g,t) effects parameters, under the conditional and unconditional parallel trends assumptions, with clustering at the municipality level. For DiDiD, the table reports the difference between aggregated ATT(g,t) effects parameters (DiD) estimated on the sample of women and men. For presentation purposes, we multiplied binary outcomes by 100. Unconditional group-time estimates were obtained following the estimator of Callaway and Sant'Anna (2021). Conditional group-time estimates in addition apply the regression-outcome DiD estimator of Sant'Anna and Zhao (2020) with pre-treatment levels of urbanization, primary-school expenditures, sex ratio of the birth cohort, and share of working women in the municipality as controls. To estimate models at the level of variation, we collapsed our data at the municipality-by-birth cohort cells with weights totaling the number of individuals in each cell. Inference procedures used bootstrapped standard errors adjusted for multiple hypothesis testing and clustering at the municipality level. *** p<0.01, ** p<0.05, * p<0.1

**For Online Publication**

**Appendices**

to the paper “Life-Cycle Effects of Comprehensive Sex Education”

by Volha Lazuka and Annika Elwert

## Appendix A – Additional Details on the Sex Education Instructions According to the 169 Royal Decree

On April 10, 1942, the Swedish Parliament issued a royal decree Nr 169 that mandated the implementation of sex education in primary schools. But the decree also stipulated that in case the class teacher did not feel capable of providing sex education the school board could assign another teacher or a medical doctor to give sex education classes, and in case no such teacher could be found, sex education was not to be taught.

Following the 1942 decree, the authorities published instructions for teaching sex education in school—the guidelines in sex education in 1942 (Ecklesiastiskdepartementet 1944) and an instruction booklet as a separate book in 1945 (Royal Board of Education in Sweden 1945). Given the novelty of the subject and the potential reticence by teachers, the booklet provided a detailed account of the reasons for introducing sex education in school and the teaching methods, including sample lectures and answers to students' potential questions. The booklet remained unchanged until 1955, the year of the law on compulsory sex education, and even then, the new version refreshed the motivation but left the curriculum and the instructions untouched.

In the primary grades (7 to 9 years old), the class on Local Geography and Nature separated students by sex and taught them about human physiology, hygiene, and the differences between the sexes. Starting from grade four to six in elementary school (ages 11–13 years), the course content included a variety of sex-related topics in the Biology and Religious Knowledge subjects, such as abstinence and masturbation, pregnancy and labor, sexual abuse, and the basics of family formation. In the final grades of elementary school (ages 14–15 years), sex education widened to include topics of sexually transmitted diseases and (low-prevention) contraception in the Biology subject. The History and Civics subject of these grades also included an extensive module on the social norms and public institutions governing sex and family life. The part on the moral and social aspects of sex was repeated in the curriculum for students who attended the ninth grade of the comprehensive school or continued their education in a secondary school (aged 16 years and older), but had no major differences compared with the earlier grades.

"Love–marriage–sex/children" was translated as the universal formula, in which persons were required to think rationally about having children, particularly in terms of costs. Young people were taught to understand the risk of "children being born without the parents having a home to offer them" (Royal Board of Education 1964, 12). In this period, when contraceptives became legalized but remained inefficacious (the first powerful contraceptive pill was introduced in 1964), a great emphasis in avoiding unintended pregnancies was put on practicing abstinence during adolescence. Importantly, the need for abstinence was motivated purely by rational choice, not moral or religious standards. To illustrate, young people were taught that unmarried parents were not to be judged morally but viewed as an example of the difficulties that pre-marital sexual intercourse could lead to in life:

> *Children born outside marriage have the same social rights as other children… It should be explained that unmarried people, especially if they have children, have special difficulties and problems, which do not exist for complete families with husband, wife, and children. This problem is dealt with in detail, but exaggerated attention should not be given to divorced people, widows, widowers, or unmarried mothers… An unmarried father or mother who does everything possible to overcome these difficulties shows a worthy public spirit, as also do those who give their help to such parents and children.* (Royal Board of Education 1964, 15)

Sex instruction for students aged 11–13 years also provided pictures with information on the development of the fetus and the course of labor, and for students aged 14–15 years, explained the welfare support for childbirth and care. Girls were recommended to observe two- to three-year intervals between births to prevent being worn out in adulthood and abortion of unwanted pregnancies. Teachers revealed the unwanted costs of sexual activity, such as sexual abuse, undesired pregnancies, and the spread of sexually transmitted diseases:

> *Teaching should aim at showing which things are not morally desirable from a personal and social point of view, and which cannot, therefore, be accepted as norms of behavior. […] Sex instruction must give them a clear understanding of the obligations which accompany sexual relations between a man and a woman. It is necessary to bring the understanding to the home that these are not just a personal matter between a man and a woman, but that they have consequences—especially for children—and are in general a social matter of great importance.* (Royal Board of Education 1964, 11–12)

In the spirit of its times, the 1942/1945 instructions informed about racial hygiene, encouraging students to prevent the birth of children "with hereditary predispositions to physical and mental illness" (Kungl. Skolöverstyrelsen 1945). Only the 1956 version excluded any mention of heritability aspects (Royal Board of Education in Sweden 1956). From 1949, sex instruction explained the employees' protection law, according to which wage-earning women could take a parental leave without losing their job or income.

Teaching on the laws and institutions regulating citizens' sexual life was revolutionary, deemed to exemplify the functioning of a democratic society:

*It is a requirement of society that children leave school with enough knowledge to understand what society demands of them as citizens. They should have learnt through the school the broad outlines of the norms which govern the way people live together in a civilized community, and the consequences of following and not following them. This is an important part of the school's educational activity*. (Royal Board of Education 1964, p. 11–12)

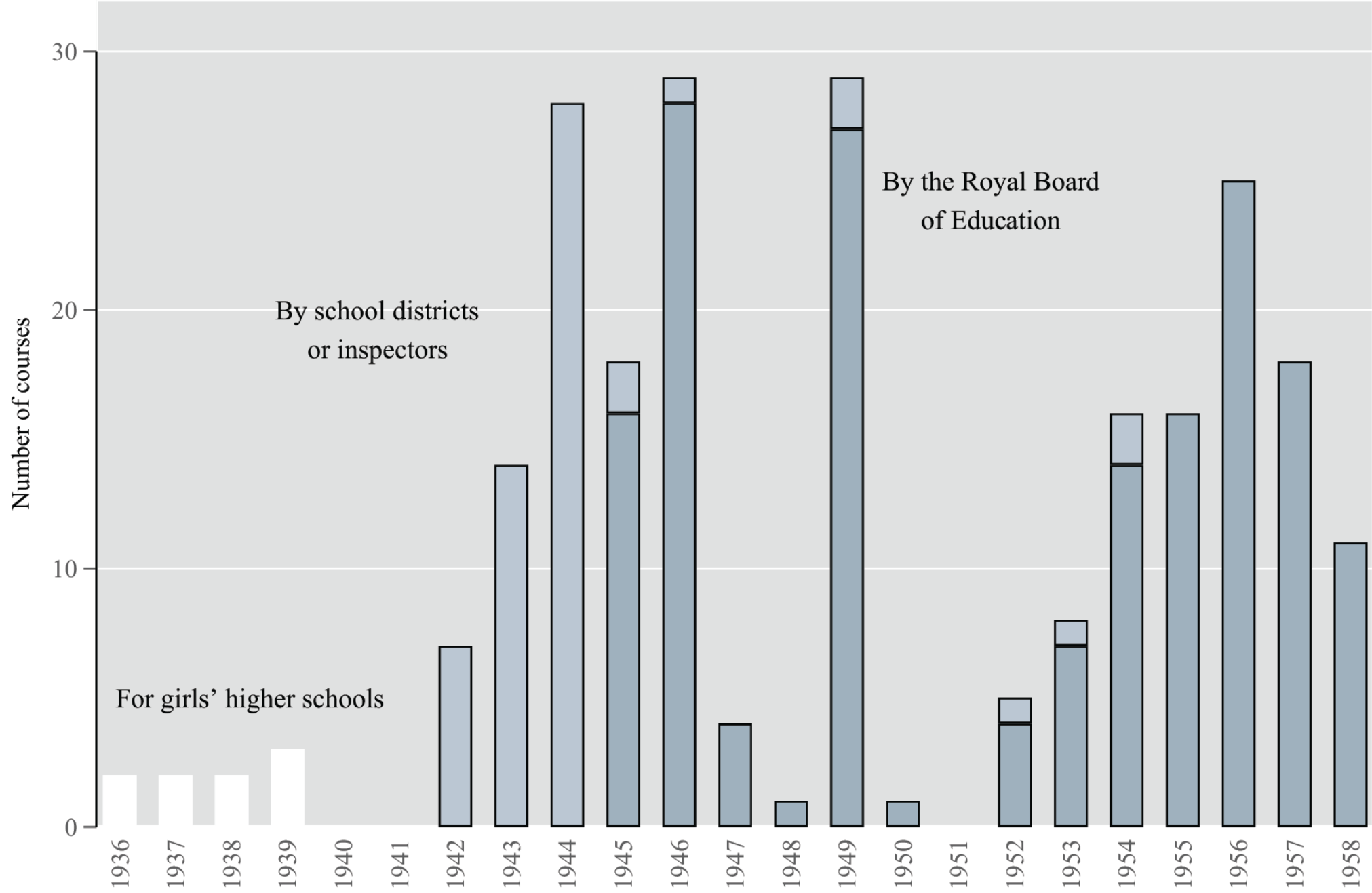

Figure A.1 – Number of teacher training courses during the sex education reform, in total per year.

*Source*: Data on reform implementation gathered from the national archives.

*Note:* Several teacher courses had already been held in large cities in 1936, but these courses had limited slots, focused on sex and racial biology, and only accepted teachers of girls' higher schools as participants (Justitiedepartementet 1951). These earlier courses played a minor role in the expansion of sex education because only 4% of girls attended upper secondary schools (Ekstedt 1976).

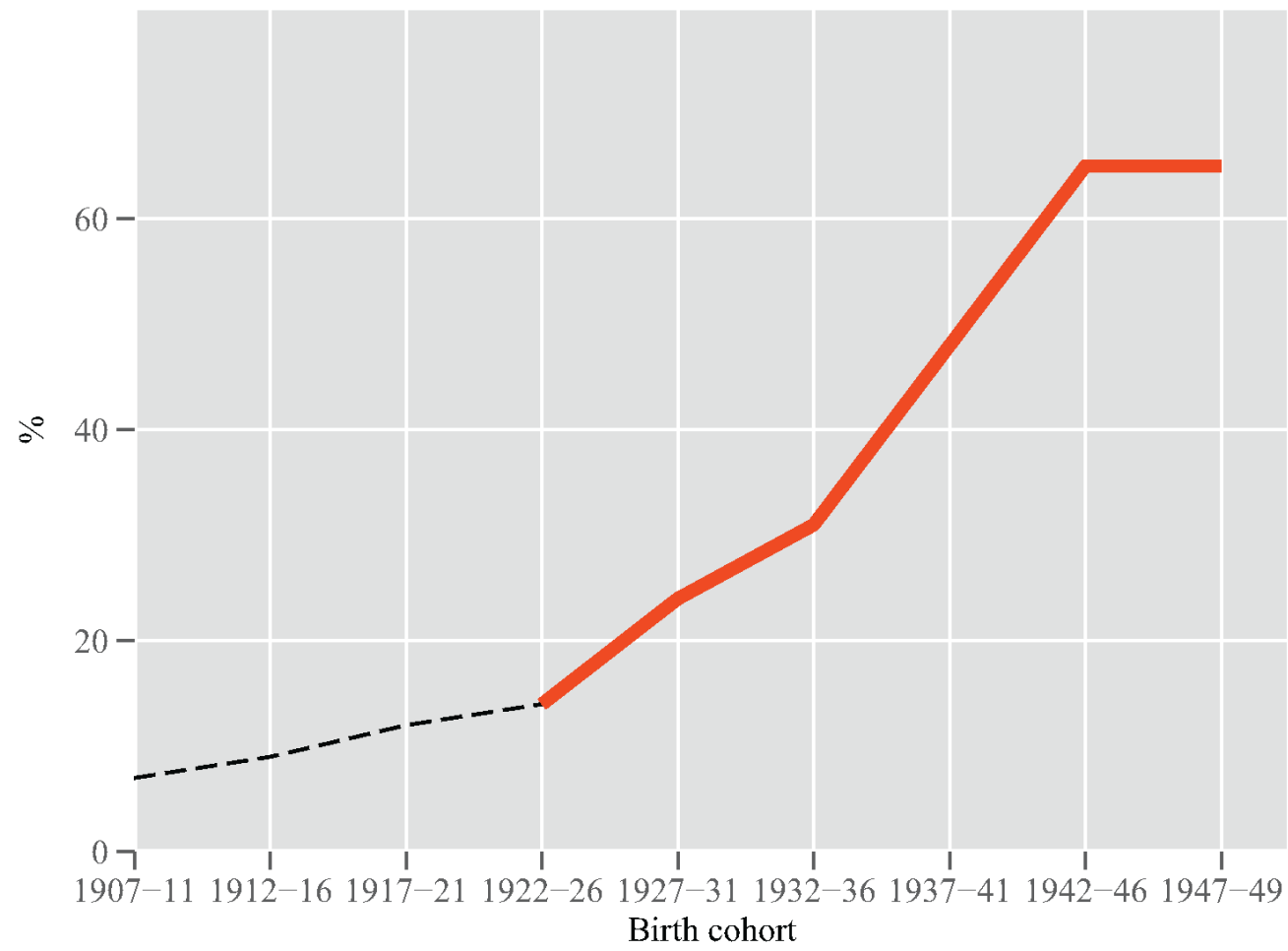

Figure A.2 – Coverage by sex education classes in primary school, aggregated by cohort born in 1907–1949.

*Source*: Calculations based on the national survey on sex habits among Swedish adults conducted in 1967 (Zetterberg 1969). The survey used a random sample of about 2,000 men and women (with a non-response rate of 9.5%), representative of the Swedish population.

## Appendix B – Theoretical Model of Utility-Reducing Sex Education

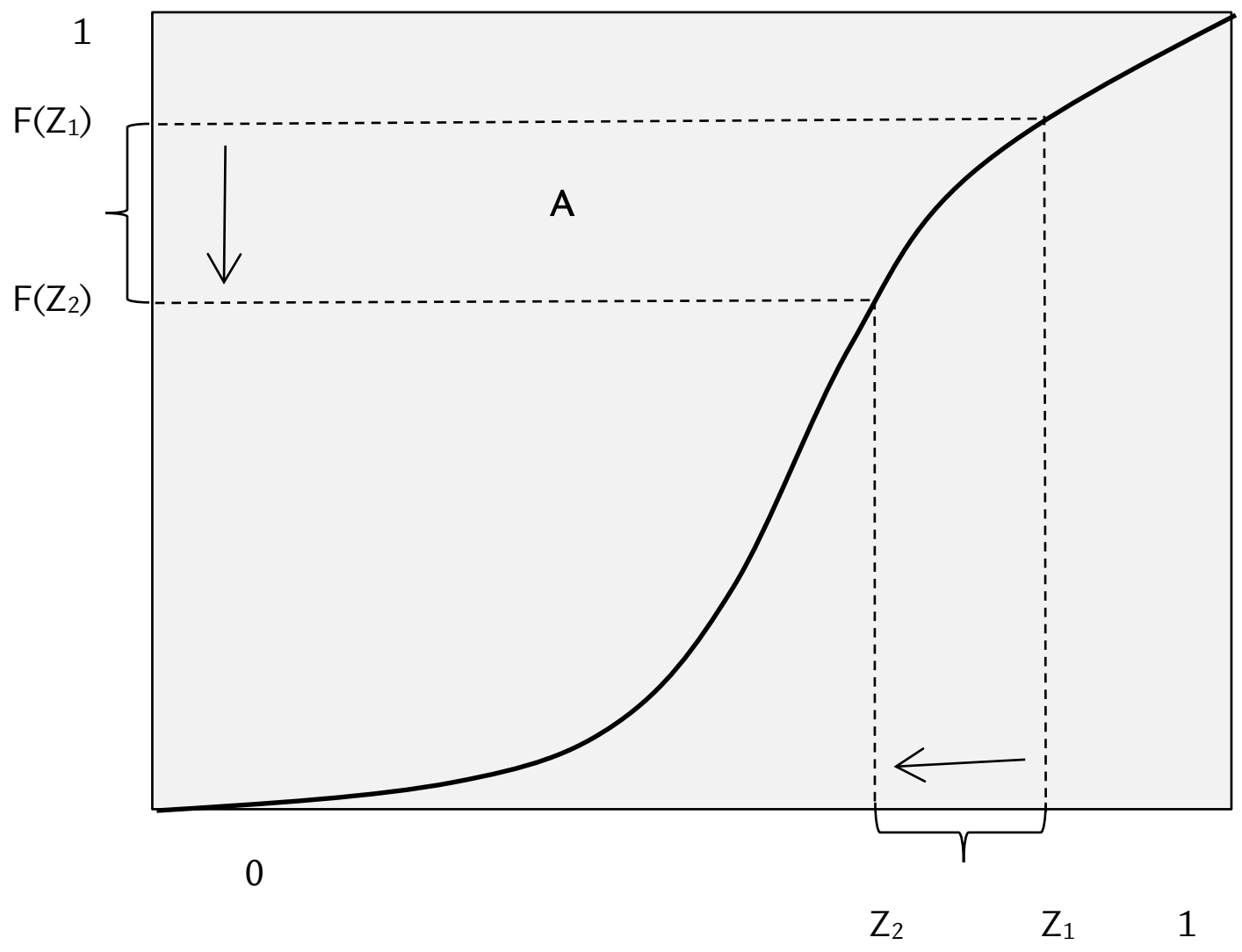

Figure B.1 – Reduction in sexual activity with utility-reducing sex education

*Source*: Own illustration based on Oettinger (1999).

*Note*: *The* new information provided was expected to alter the perceived costs and/or benefits of sexual activity, as proposed by Oettinger (1999). Individuals could then make the informed decision to engage in sexual activity (as opposed to abstaining) based on a cost-benefit analysis in which the perceived benefits outweighed the perceived costs. These benefits included immediate gratification, the development of an ongoing relationship, an increase in social status among peers, or the acquisition of knowledge. Meanwhile, the inherent costs associated with sexual activity included the deterioration of an ongoing relationship, disapproval from parents or society, and the possibility of illness or undesired pregnancy. The greater the amount of information provided by sex education, the more significant the potential impact on a teenager's sexual behavior.

The utility-*altering* sex education that reduced the utility of sexual activity was expected to ultimately reduce the number of sexually active individuals and unintended pregnancies. Such sex education, for example, revealed the benefits of abstinence or the perceived individual and societal costs of sexual activity. These components were implemented by the sex education reformers in 1942–1958 in Sweden. Abstinence was taught to be a social norm and a source of personality growth, and unintended pregnancies and venereal diseases were posited as immense costs for pupils themselves, their families, and society. Moreover, teachers were instructed to by no means encourage teens to engage in any sexual activity.

## Appendix C – Results of the Tests for Negative Weights and Heterogeneous Effects in a Two-Way Fixed-Effects Regression

In the case of differential treatment timing in the reform implementation, a presence of both negative weights and heterogeneous treatment effects leads to a bias in the DiD estimate, if the estimate is obtained from a two-way fixed-effects regression (Roth et al. 2023; Clement de Chaisemartin and D'Haultfoeuille 2022; Goodman-Bacon 2021; Jakiela 2021). We expect that negative weights appear in our sample: a substantial portion of municipalities are treated in the first years of the reform implementation. Below we provide the results of the tests for a presence of both negative weights and heterogeneous treatment effects, based on the marriage of convenience by age 18 as an outcome. We aggregate the sample by year of birth X municipality cells and weight observations by a number of individuals in each cell.

The two-way fixed-effects OLS estimator provides a sex education reform estimate equal to -0.288 (standard error is 0.083), a highly statistically and economically significant effect. However, we first detect that the treatment effects are heterogenous between combinations of treatment and comparison groups. A Goodman-Bacon (2021)'s decomposition shows that an estimate for the comparison of the ever- to never-treated municipalities amounts to -0.172 [negative weights cannot appear in this group], while it is -0.720 for the comparison of later- to earlier-treated municipalities, a source of negative weights. Therefore, an estimate from a subgroup of municipalities, which create so-called "forbidden" comparison in DD, suggests a much larger and potentially biased reform effect.

Second, following Clément de Chaisemartin and D'Haultfœuille (2020) and Jakiela (2021), we calculate weights as residuals from a regression of treatment on two-way fixed effects and plot them as fractions in Figure C.1. As can be seen, a fraction of municipality-year of birth observations with negative weights is unbalanced between treatment groups. In total, observations for 358,336 individuals who were treated receive a negative weight, 25 percent of the whole sample.

The presence of both such weights and heterogenous treatment effects implies that a two-way fixed-effects OLS estimator provides an DiD estimate that differ from the one representing an average treatment effect on the treated. In the case of marriage for convenience, the effect is biased downwards.

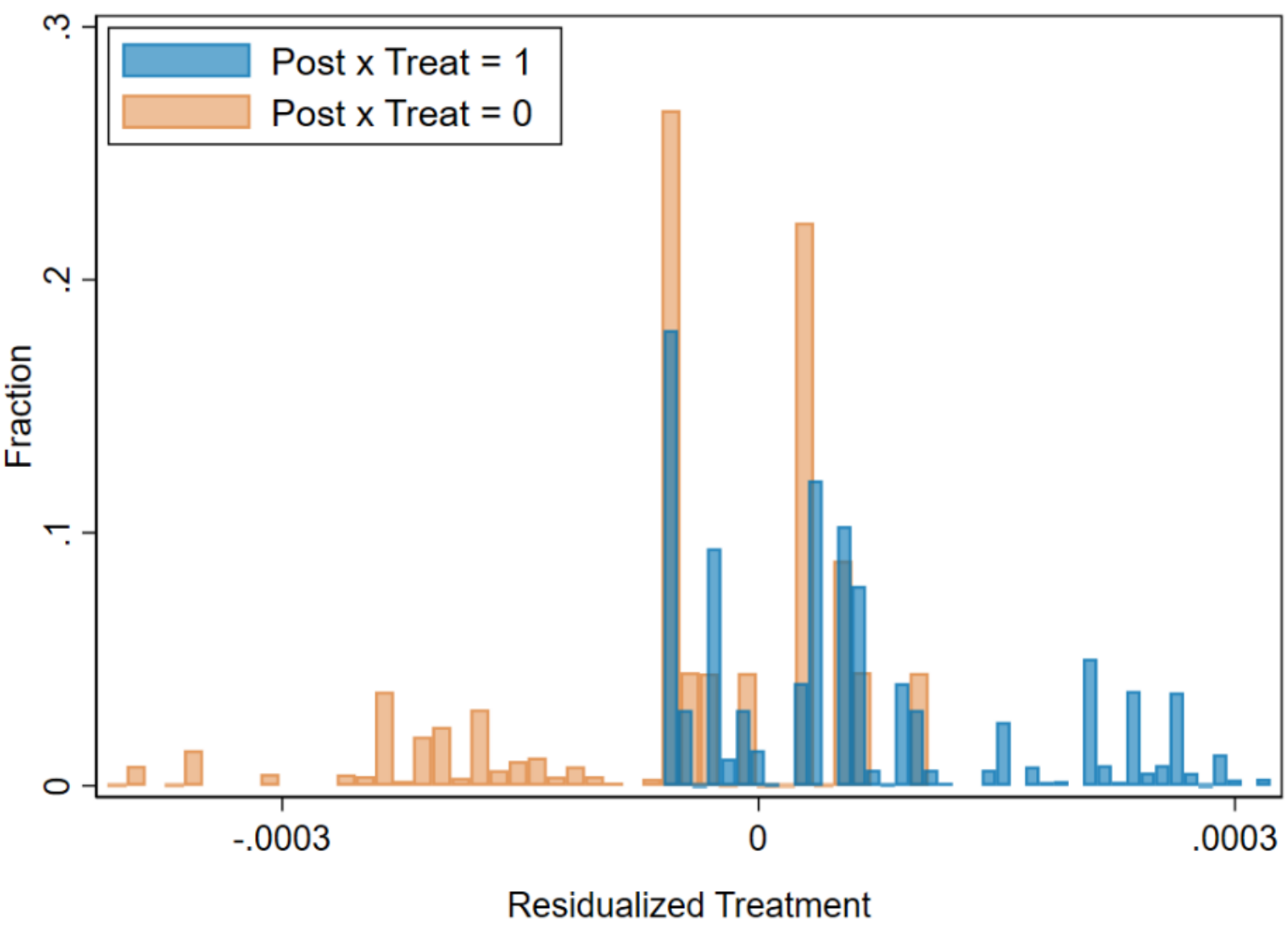

Figure C.1 — Two-way fixed-effects weights, by treatment status.

*Note*: The figure presents a histogram of weights used to calculate the two-way fixed-effects estimates of the impact of sex education on marriage of convenience by age 18. The weights are the residuals from a regression if treatment on municipality and year of birth fixed effects, scaled by the sum of squared residuals across all observations.

## Appendix D – Determinants of the Sex Education Reform Timing

| | **(A) Years 1942–1958** | **(B) Years 1946–1958** |
|---|---|---|
| | **Treatment year** | **Treatment year** |
| | (1) | (2) |
| | | |
| Share of unemployed women in 1940 | 5.138*** | 1.072 |
| | (0.469) | (0.708) |
| School investments in 1940 | -1.135** | -0.353 |
| | (0.538) | (0.515) |
| Population in 1940 | -0.225 | -0.862 |
| | (0.730) | (0.602) |
| Share of votes for Social Democrats in 1940 | -1.595** | -0.587 |
| | (0.771) | (0.657) |
| Sex ratio of 1930's birth cohort | 0.058 | 0.119 |
| | (0.175) | (0.151) |
| | | |
| Mean of dep.var. | 1953.304 | 1956.159 |
| Observations | 2,503 | 1,939 |
| R-squared | 0.069 | 0.070 |

*Note:* Table reports the estimates of the municipality-level covariates in 1940 (binary variables divided at the median) for continuous treatment initiation ("Year of treatment", with never-treated set to 1959). Standard errors are in parentheses.

*** p<0.01, ** p<0.05, * p<0.1

# Appendix E – Additional Analysis

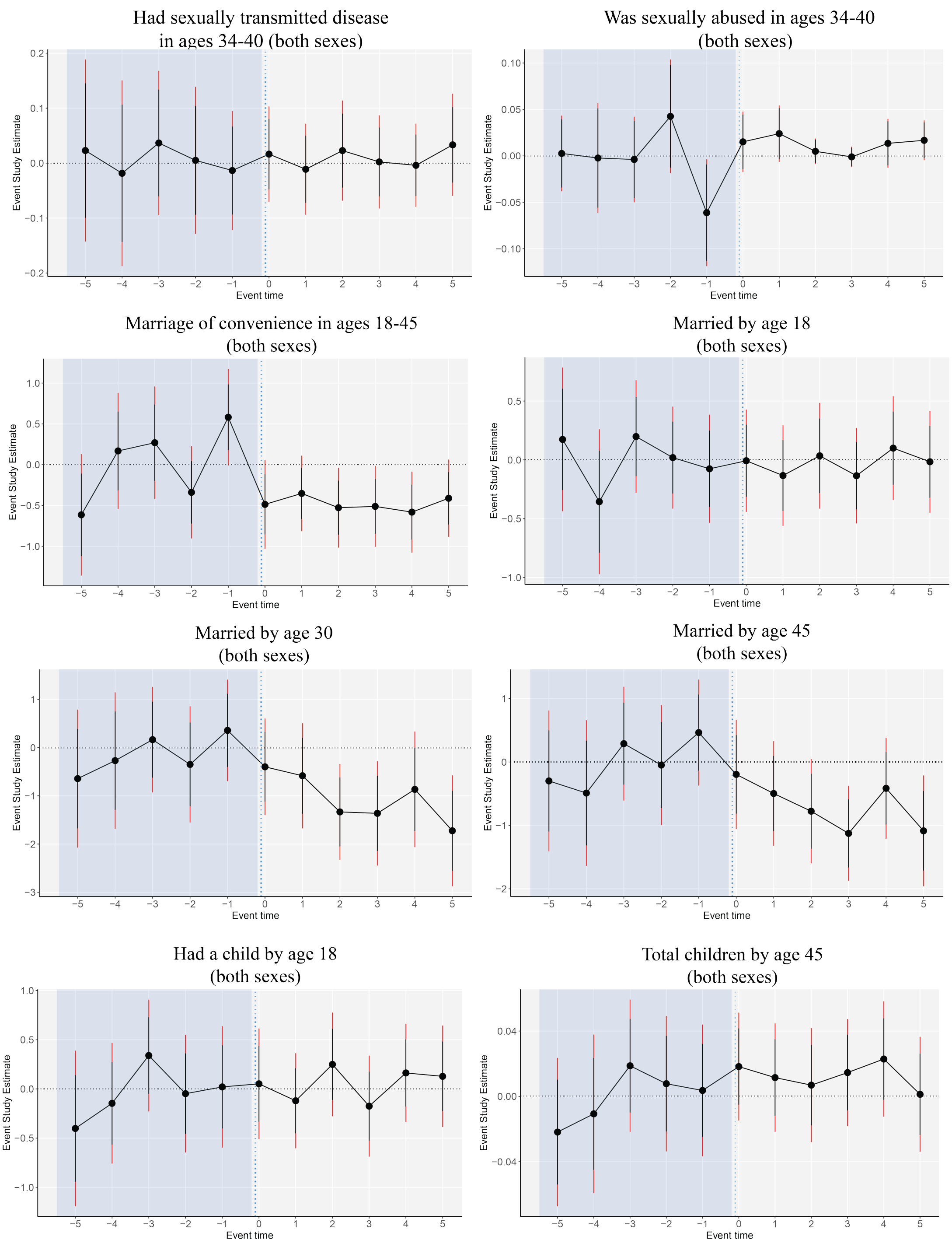

Figure E.1 – Event studies for sexual activity and family formation, ages 18–45 years (not shown in the main body of the paper).

*Note*: The figure reports ATT(g,t) effects parameters aggregated by event time and weighted by group size, under the unconditional parallel trends assumption. To ease interpretation, we multiplied binary outcomes by 100. The inference procedures used both asymptotic and bootstrapped standard errors (with wider confidence intervals). Both types of standard errors adjust for clustering at the municipality level, while bootstrapped standard errors also adjust for multiple hypothesis testing.

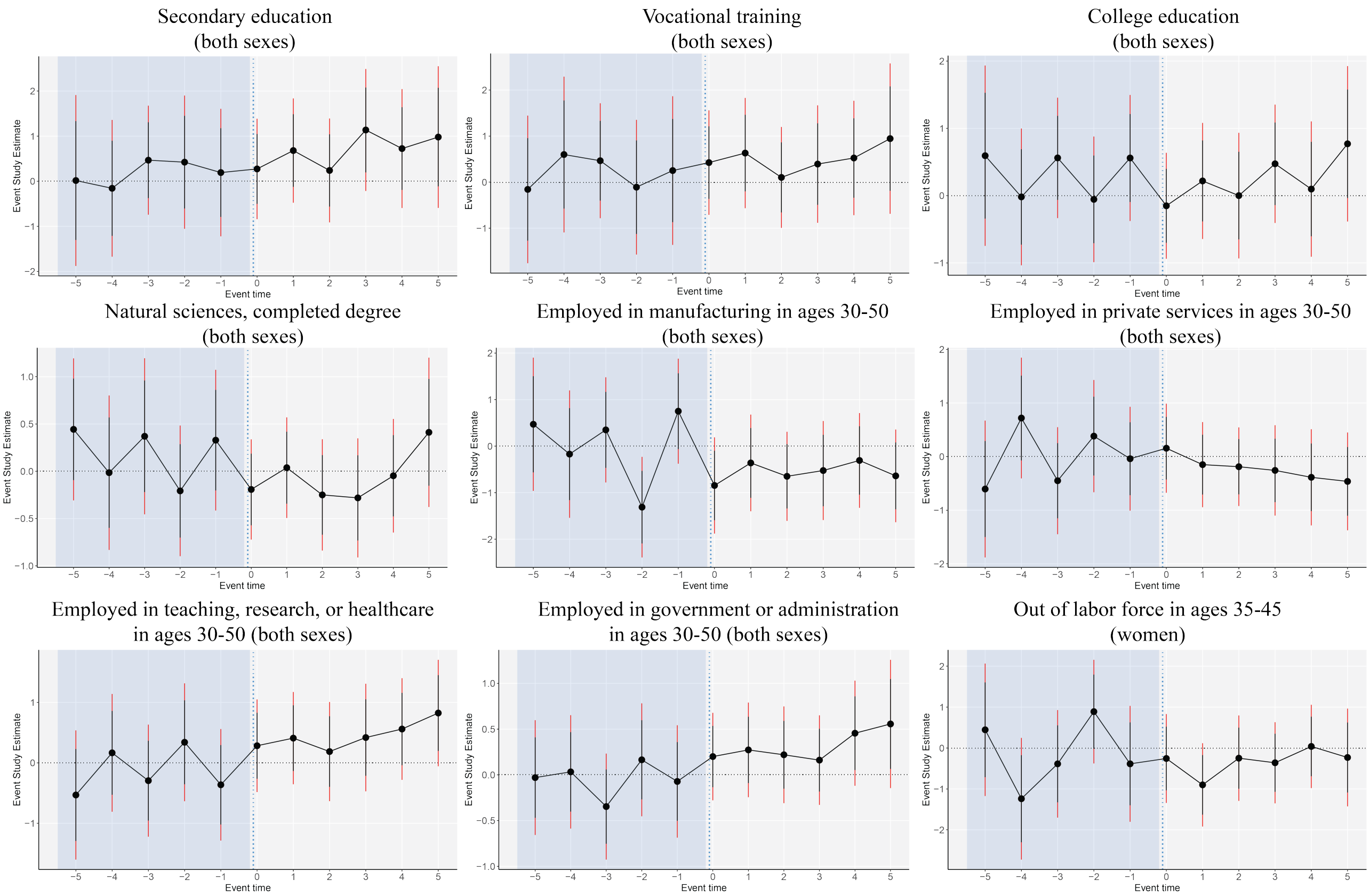

Figure E.2 – Event studies for educational and labor market outcomes (not shown in the main body of the paper).

*Note*: The figure reports ATT(g,t) effects parameters aggregated by event time and weighted by group size, under the unconditional parallel trends assumption. To ease interpretation, we multiplied binary outcomes by 100. The inference procedures used both asymptotic and bootstrapped standard errors (with wider confidence intervals). Both types of standard errors adjust for clustering at the municipality level, while bootstrapped standard errors also adjust for multiple hypothesis testing.

## Appendix F – Robustness Analysis: The Validity of the Estimator and Interaction with Other Social Reforms

In this Appendix, we perform (1) estimations with alternative control groups and (2) with alternative DiD estimators. (3) We also analyze the interactions of the sex education reform with other reforms. Our results are robust to the use of an alternative control group and the use of other estimators for heterogeneous treatment effects (i.e., imputation estimator).

(1) First, we perform group-time DiD estimations using a group of not-yet-treated municipalities as a control. In our baseline estimations, we only include never-treated municipalities ($S_g$ = 0) into the comparison group, because namely the comparison to this group would require a "weak" parallel trend assumption—parallel development only for one year prior to the treatment. In contrast, when not-yet-treated are included into the comparison group, parallel-trends must hold for longer pre-treatment periods because not-yet-treated participate in both a control group (before treatment) and a treated group. However, our results are not be sensitive if we additionally include not-yet-treated municipalities or use only them as a comparison

For difference-in-differences (DiD), tables F.1–F.6 below report aggregated (average, weighted by group size) ATT(g,t) effects parameters, under the conditional and unconditional parallel trends assumptions, with clustering at the municipality level. For difference-in-difference-in-differences (DiDiD), the tables below report the difference between aggregated ATT(g,t) effects parameters (DiD) estimated on the sample of women and men. For presentation purposes, we multiplied binary outcomes by 100. Unconditional group-time estimates were obtained following the estimator of Callaway and Sant'Anna (2021). Conditional group-time estimates in addition apply the regression-outcome DiD estimator of Sant'Anna and Zhao (2020) with pre-treatment levels of urbanization, primary-school expenditures, sex ratio of the birth cohort, and share of working women in the municipality as controls. Inference procedures used bootstrapped standard errors.

(2) Second, we apply an imputation estimator Borusyak, Jaravel, and Spiess (2024), which offers a valid alternative to the group-time estimator. Other imputation approaches are analogous (Gardner 2022; Liu, Wang, and Xu 2022). The imputation estimator fits a TWFE regression with year and unit fixed effects only for the not-yet-treated units, infers the never-treated potential outcome for each treated unit using the predicted value from that TWFE regression, and then aggregates the unit-specific ATT estimates into the summary ATT parameter with similar weights as in the group-time estimator (i.e, group sizes). The estimator relies on the "long" parallel trends (i.e., for all pre- and post-treatment periods) and no anticipation assumptions. While the group-time estimator makes all comparisons relative to the last pre-treatment period, the imputation estimator makes comparisons relative to the average of the pre-treatment periods (see Roth et al. 2023 for more details).

In Table F.7, we perform estimations for women's sexual activity outcomes based on the imputation approach and obtain the estimates nearly identical to the group-time estimates. Additionally, we present the estimates for the pre-treatment years, as a test for parallel trends. As in the case of group-time estimator, pre-treatment estimates based on the imputation approach are never statistically significant. Finally, in Table F.8, we show the TWFE estimates for informative purposes—TWFE estimates are severely biased.

(3) Table F.8 and Figure F.1 present the aggregated (DiD) and event-study (DiDiD) estimates for the interaction effects of the childbirth reform with the sex education reform for the marriage of convenience in ages 18–45 years. The results for the other outcomes provide a similar pattern. Because we had to exclude municipalities for which the assignment of treatment by social reforms was imprecise (as in Lazuka 2023), we first ascertain that the sex education reform effects remained unchanged for the restricted estimation sample. We find that the sex education reform effects do not differ economically and statistically between the subsamples with municipalities that introduced another reform and those that never or had not done so. In relation to the childbirth reform, the absence of interaction effects suggests that maternal empowerment, which was the reform's target, did not affect the first generation's outcomes.

Arguably, two other events could have influenced the estimates of the sex education reform. First, starting in the 1940s, the authorities gradually extended mandatory schooling to 9 years (Holmlund 2020). In the context of our study design, this schooling reform implied that cohorts at age 16 years, which we considered untreated, could potentially have received sex education. However, our analysis revealed that only 3.9% of students within our cohorts were affected by the schooling reform. This minimal impact suggested that any potential bias on the effects would be negligible. Additionally, our results from the event studies did not provide any evidence of anticipatory effects, which would manifest as a spike in the first year before the reform. Second, in other European countries that participated in WWII, adult women (i.e., the parental generation) experienced favorable labor market conditions in the 1940s–1950s owing to wartime casualties among men. Notably, Sweden did not participate in WWII. In summary, our analysis indicated that no other events influenced the outcomes of the sex education reform.

Table F.1 – Effects on sexual activity and abuse, ages 34–40 years.

| | **Abortion (women)** | **Cervical cancer (women)** | **Sexually transmitted diseases (both)** | **Sexual abuse (both)** |
|---|---|---|---|---|
| **(a) Unconditional DiD** | -1.085*** | -1.012*** | 0.004 | 0.004 |
| | (0.273) | (0.393) | (0.027) | (0.003) |
| **DiDiD, women minus men** | | | -0.017 | -0.010 |
| | | | (0.040) | (0.011) |
| | | | | |
| **(b) Conditional DiD** | -1.092*** | -1.007*** | 0.008 | 0.001 |
| | (0.250) | (0.404) | (0.029) | (0.004) |
| **DiDiD, women minus men** | | | -0.013 | -0.013 |
| | | | (0.041) | (0.014) |
| | | | | |
| **Pre-mean** | 1.358 | 4.342 | 0.053 | 0.017 |
| **Individuals** | 379,494 | 379,494 | 776,930 | 776,930 |

*** p<0.01, ** p<0.05, * p<0.1

Table F.2 – Effects on marriage, ages 18–45 years.

| | **Marriage of convenience (both)** | | **Married (both)** | | | |
|---|---|---|---|---|---|---|
| | by 18 | 18-45 | by 18 | by 25 | by 30 | by 45 |
| **(a) Unconditional DiD** | -0.168*** | -0.385*** | -0.041 | -2.022*** | -1.634*** | -0.839*** |
| | (0.064) | (0.147) | (0.126) | (0.476) | (0.412) | (0.253) |
| **DiDiD, women minus men** | -0.291*** | -0.084 | 0.015 | 0.583 | 0.315 | 0.175 |
| | (0.100) | (0.225) | (0.197) | (0.601) | (0.518) | (0.390) |
| | | | | | | |
| **(b) Conditional DiD** | -0.165*** | -0.307** | 0.048 | -1.885*** | -1.295*** | -0.637** |
| | (0.064) | (0.148) | (0.130) | (0.477) | (0.389) | (0.256) |
| **DiDiD, women minus men** | -0.261** | -0.121 | 0.174 | 0.386 | -0.331 | -0.162 |
| | (0.104) | (0.244) | (0.198) | (0.656) | (0.575) | (0.427) |
| | | | | | | |
| **Pre-mean** | 0.608 | 3.534 | 2.366 | 54.741 | 77.176 | 86.904 |
| **Individuals** | 1,026,358 | 989,655 | 1,026,358 | 1,020,490 | 1,015,396 | 989,655 |

*** p<0.01, ** p<0.05, * p<0.1

Table F.3 – Effects on parenthood, ages 18–45 years.

| | **Having a child (both)** | | | | **Total number of children (both)** |
|---|---|---|---|---|---|
| | by 18 | 18-25 | 25-30 | 30-45 | |
| **(a) Unconditional DiD** | 0.045 | -1.328*** | -0.361 | 1.707*** | 0.021* |
| | (0.144) | (0.488) | (0.392) | (0.414) | (0.012) |
| **DiDiD, women minus men** | -0.030 | 0.776 | -0.662 | 0.709 | 0.017 |
| | (0.225) | (0.607) | (0.581) | (0.583) | (0.015) |
| | | | | | |
| **(b) Conditional DiD** | 0.061 | -0.957*** | -0.395 | 1.731*** | 0.025** |
| | (0.149) | (0.436) | (0.412) | (0.425) | (0.012) |
| **DiDiD, women minus men** | 0.091 | 0.233 | -1.056 | 1.253** | 0.020 |
| | (0.237) | (0.619) | (0.637) | (0.650) | (0.016) |
| | | | | | |
| **Pre-mean** | 3.462 | 50.188 | 45.277 | 43.242 | 2.008 |
| **Individuals** | 1,026,358 | 1,020,490 | 1,015,396 | 989,655 | 989,655 |

*** p<0.01, ** p<0.05, * p<0.1

Table F.4 – Effects on completed education.

| | **Years of schooling (both)** | **Secondary (both)** | **Vocational training (both)** | **College (both)** | **Social sciences, law, and administration (both)** | **Natural sciences (both)** |
|---|---|---|---|---|---|---|
| **(a) Unconditional DiD** | 0.036* | 0.739* | 0.837** | 0.480 | 0.890*** | 0.049 |
| | (0.021) | (0.426) | (0.431) | (0.394) | (0.331) | (0.205) |
| **DiDiD, women minus men** | -0.010 | 0.344 | 0.085 | -0.527 | 0.432 | 0.142 |
| | (0.026) | (0.589) | (0.664) | (0.499) | (0.470) | (0.377) |
| | | | | | | |
| **(b) Conditional DiD** | 0.033** | 0.664 | 0.627 | 0.335 | 0.905*** | -0.136 |
| | (0.017) | (0.442) | (0.461) | (0.332) | (0.297) | (0.207) |
| **DiDiD, women minus men** | -0.029 | -0.211 | -0.482 | -0.826 | -0.058 | 0.357 |
| | (0.027) | (0.655) | (0.737) | (0.506) | (0.505) | (0.391) |
| | | | | | | |
| **Pre-mean** | 7.933 | 19.209 | 19.496 | 8.417 | 9.954 | 4.301 |
| **Individuals** | 969,247 | 969,247 | 969,247 | 969,247 | 969,247 | 969,247 |

*** p<0.01, ** p<0.05, * p<0.1

Table F.5 – Effects on sector of employment and occupation, ages 30–50 years.

| | **Sector of employment (both)** | | | | | **Occupational score (both)** |
|---|---|---|---|---|---|---|
| | Manufacturing | Private services | Public services | Teaching, research, and healthcare | Government and administration | |
| **(a) Unconditional DiD** | -0.570* | -0.403* | 0.761*** | 0.479** | 0.457*** | 0.286*** |
| | (0.311) | (0.229) | (0.321) | (0.247) | (0.169) | (0.084) |
| **DiDiD, women minus men** | 1.116** | -0.031 | -0.399 | 0.229 | -0.272 | 0.152 |
| | (0.464) | (0.388) | (0.467) | (0.353) | (0.295) | (0.148) |
| | | | | | | |
| **(b) Conditional DiD** | -0.282 | -0.462* | 0.723*** | 0.419 | 0.327** | 0.286*** |
| | (0.334) | (0.244) | (0.362) | (0.268) | (0.152) | (0.099) |
| **DiDiD, women minus men** | 0.785 | -0.189 | -0.773 | 0.344 | -0.199 | -0.103 |
| | (0.471) | (0.410) | (0.512) | (0.384) | (0.333) | (0.169) |
| | | | | | | |
| **Pre-mean** | 28.271 | 16.269 | 28.879 | 17.105 | 14.579 | 58.145 |
| **Individuals** | 1,010,713 | 1,010,713 | 1,010,713 | 1,010,713 | 1,010,713 | 1,006,639 |

*** p<0.01, ** p<0.05, * p<0.1

Table F.6 – Effects on labour market outcomes.

| | **Log earnings (both sexes)** | **Out of labor force (women)** | | | **In household production (women)** |
|---|---|---|---|---|---|
| | ages 48-50 | ages 30-35 | ages 35-45 | ages 45-55 | ages 35-45 |
| **(a) Unconditional DiD** | 0.104*** | -1.528*** | -0.466*** | -0.600*** | -1.350*** |
| | (0.046) | (0.387) | (0.227) | (0.211) | (0.256) |
| **DiDiD, women minus men** | 0.059 | | | | |
| | (0.068) | | | | |
| | | | | | |
| **(b) Conditional DiD** | 0.125*** | -1.366*** | -0.591*** | -0.627*** | -1.048*** |
| | (0.046) | (0.393) | (0.242) | (0.247) | (0.216) |
| **DiDiD, women minus men** | 0.040 | | | | |
| | (0.070) | | | | |
| | | | | | |
| **Pre-mean** | 10.684 | 31.881 | 18.349 | 15.903 | 10.318 |
| **Individuals** | 981,934 | 492,074 | 490,729 | 483,308 | 397,757 |

Table F.7 – Effects on sexual activity (women) with different estimators, ages 34–40 years.

| | Abortion (women) | | | Cervical cancer (women) | | |
|---|---|---|---|---|---|---|
| | TWFE | Group-time | Imputation | TWFE | Group-time | Imputation |
| **(a) Unconditional DiD** | -0.293*** | *-1.061**** | -1.091*** | -0.0721 | *-1.024**** | -1.048** |
| | (0.0997) | *(0.282)* | (0.410) | (0.156) | *(0.400)* | (0.405) |
| | | | | | | |
| **(b) Conditional DiD** | -0.305*** | *-1.079**** | -1.067*** | -0.0360 | *-0.986**** | -1.043** |
| | (0.0948) | *(0.264)* | (0.407) | (0.161) | *(0.391)* | (0.405) |
| | | | | | | |
| **(c) Pre-trends** | | | | | | |
| Event-year (-5) | -0.00181 | *-0.449* | -0.00311 | 0.00146 | *0.144* | 0.00239 |
| | (0.00192) | *(0.241)* | (0.00228) | (0.00394) | *(0.447)* | (0.00406) |
| Event-year (-4) | -0.000175 | *0.275* | 0.000545 | -0.00299 | *-0.112* | 0.000344 |
| | (0.00248) | *(0.282)* | (0.00346) | (0.00333) | *(0.497)* | (0.00368) |
| Event-year (-3) | -0.0000653 | *0.0592* | 0.000812 | 0.000329 | *0.576* | 0.00361 |
| | (0.00212) | *(0.265)* | (0.00417) | (0.00356) | *(0.470)* | (0.00381) |
| Event-year (-2) | -0.00326 | *-0.478* | -0.000752 | 0.00246 | *-0.265* | 0.00601 |
| | (0.00196) | *(0.241)* | (0.00423) | (0.00392) | *(0.454)* | (0.00444) |
| Event-year (-1) | 0.00232 | *0.368* | 0.0317 | 0.00137 | *-0.285* | 0.00133 |
| | (0.00202) | *(0.202)* | (0.00542) | (0.00332) | *(0.429)* | (0.00396) |
| | | | | | | |
| **Pre-mean** | 1.575 | 1.358 | 1.578 | 5.028 | 4.342 | 5.133 |
| **Individuals** | 588,750 | 379,494 | 106,656 | 588,750 | 379,494 | 106,656 |

*** p<0.01, ** p<0.05, * p<0.1

Table F.8 – Interaction effects with a social reform, marriage of convenience in ages 18–45.

| | **Sex education reform on a restricted sample** | **Childbirth care reform** | |
|---|---|---|---|
| | | In place | Not in place |
| **(a) Unconditional DiD** | -0.534*** | -0.506*** | -0.525* |
| | (0.172) | (0.203) | (0.310) |
| **(b) Conditional DiD** | -0.489*** | -0.472*** | -0.524* |
| | (0.170) | (0.205) | (0.314) |
| | | | |
| **Pre-mean** | 3.637 | 3.374 | 3.932 |
| **Individuals** | 881,444 | 607,520 | 273,924 |

*** p<0.01, ** p<0.05, * p<0.1

*Note*: For difference-in-differences (DiD), the table reports aggregated (average, weighted by group size) ATT(g,t) effects parameters, under the conditional and unconditional parallel trends assumptions, with clustering at the municipality level. For difference-in-difference-in-differences (DiDiD), the table reports the difference between aggregated ATT(g,t) effects parameters (DD) estimated on the sample of women and men. For presentation purposes, we multiplied binary outcomes by 100. Unconditional group-time estimates were obtained following the estimator of Callaway and Sant'Anna (2021). Conditional group-time estimates in addition apply the regression-outcome DD estimator of Sant'Anna and Zhao (2020) with pre-treatment levels of urbanization, primary-school expenditures, sex ratio of the birth cohort, and share of working women in the municipality as controls. Inference procedures used bootstrapped standard errors adjusted for multiple hypothesis testing and clustering at the municipality level.

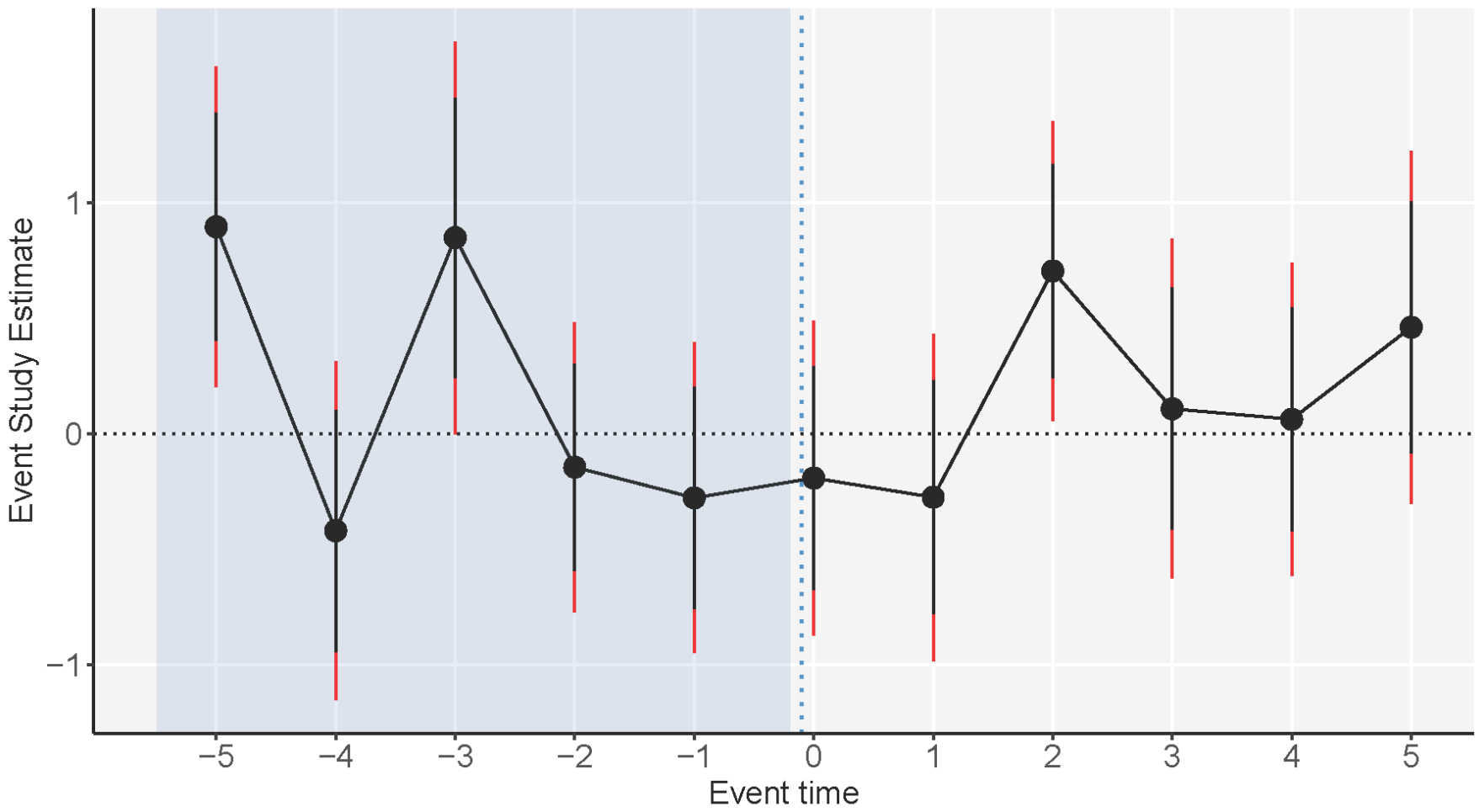

Figure F.1 – Event studies for the interaction effects of the sex education reform with the childbirth reform.

*Note*: The figure reports differences in the ATT(g,t) effects parameters (DiDiD effects) with versus without the hospital childbirth reform, aggregated by event time and weighted by group size, under the unconditional parallel trends assumption and with clustering at the municipality level. The inference procedures used both asymptotic and bootstrapped standard errors (with wider confidence intervals). Both types of standard errors adjust for clustering at the municipality level, while bootstrapped standard errors also adjust for multiple hypothesis testing.

## References in Appendices